\documentclass[ reprint,
superscriptaddress,
 amsmath,amssymb,
 aps, floatfix
]{revtex4-2}
\usepackage{graphicx}
\usepackage{dcolumn}
\usepackage{bm}
\usepackage{color}
\usepackage{amsmath}
\usepackage{amssymb}
\usepackage{braket}
\usepackage{amsfonts}
\usepackage{soul}
\usepackage{url}
\usepackage{array,tabularx}
\usepackage{graphics}
\usepackage{times}
\usepackage{subfigure}
\usepackage[export]{adjustbox}
\usepackage[mathscr]{euscript}
\newcommand\x{3.2}
\newcommand\y{8.0}
\newcommand\z{4.0}

\begin{document}

\title{Emerging exceptional point with breakdown of skin effect in non-Hermitian systems}

\author{Sayan Jana}\email{sayanjana@tauex.tau.ac.il}
\affiliation{School of Mechanical Engineering, Tel Aviv University, Tel Aviv 69978, Israel}
\author{Lea Sirota}\email{leabeilkin@tauex.tau.ac.il}
\affiliation{School of Mechanical Engineering, Tel Aviv University, Tel Aviv 69978, Israel}

\begin{abstract}

We study the interplay of two distinct non-Hermitian parameters: directional coupling and onsite gain-loss, together with topology, in coupled one-dimensional (1D) non-Hermitian Su-Schrieffer-Heeger (SSH) chains. The SSH model represents one of the simplest two-band models that features boundary localized topological modes. Our study shows how the merging of two topological modes can lead to a striking spectral feature of non-Hermitian systems, namely exceptional point (EP). We reveal the existence EP as a singularity in the parameter space of non-Hermitian couplings carrying a half-integer topological charge. We also demonstrate two different localization behaviors observed in the bulk and hybridized topological modes. While the bulk states and individual topological modes remain localized at the boundaries due to skin effect, the competition between the constituent non-Hermitian parameters can overcome the strength of skin effect and lead to the complete \textit{delocalization} of these hybridized modes. We obtain explicit analytic solutions for the eigenfunction and the eigenenergy of the hybridized modes, which exactly match the numerical results and successfully reveal the underlying cause of delocalization and the emergence of EP.

\end{abstract}

\maketitle

\begin{figure}[tb]
\begin{tabular}{c c}
 \textbf{(a)}  &   \textbf{(b)} \\
  \includegraphics[height=3cm, width=3.4cm, valign=c]{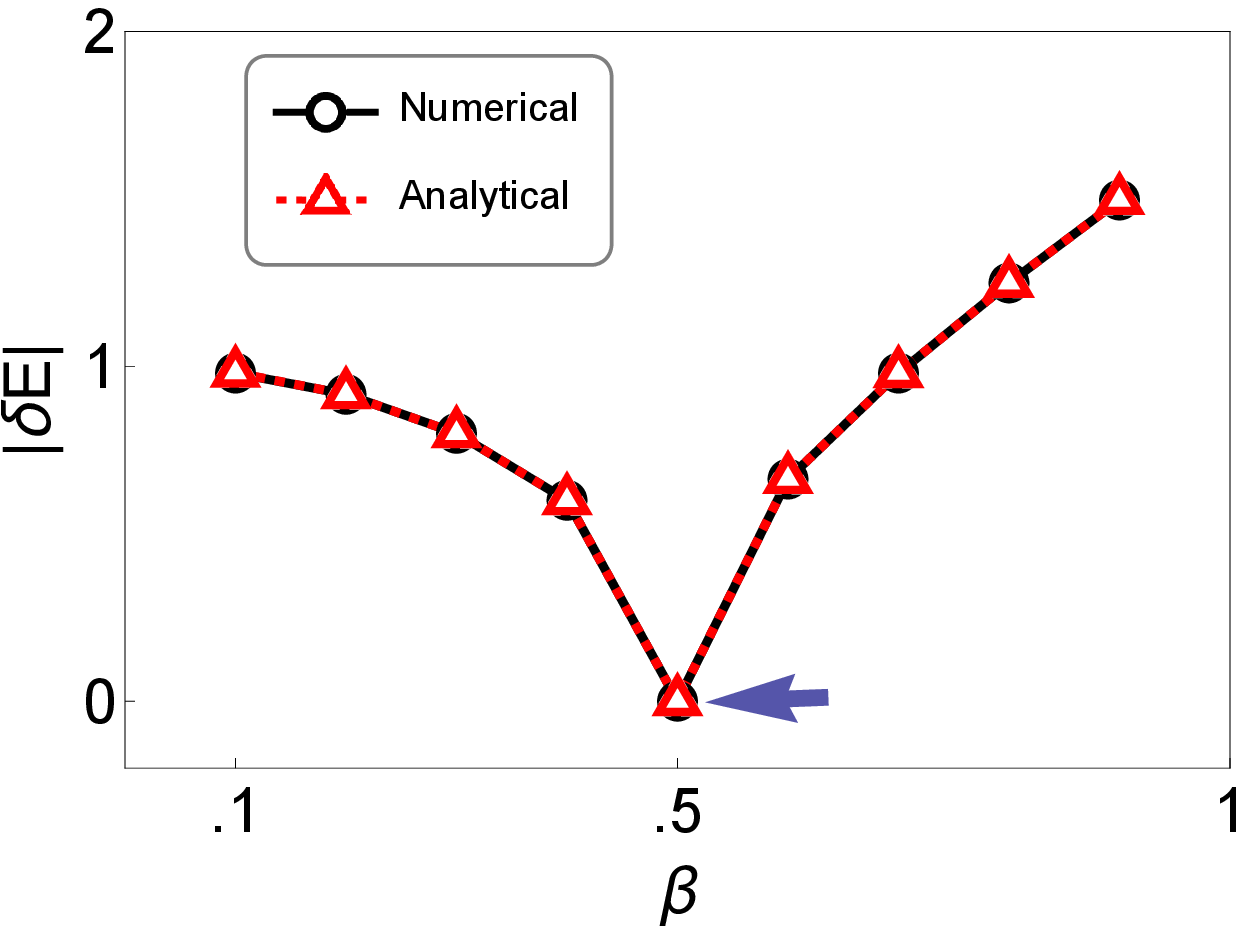}  &    \includegraphics[height=3cm, width=3.4cm, valign=c]{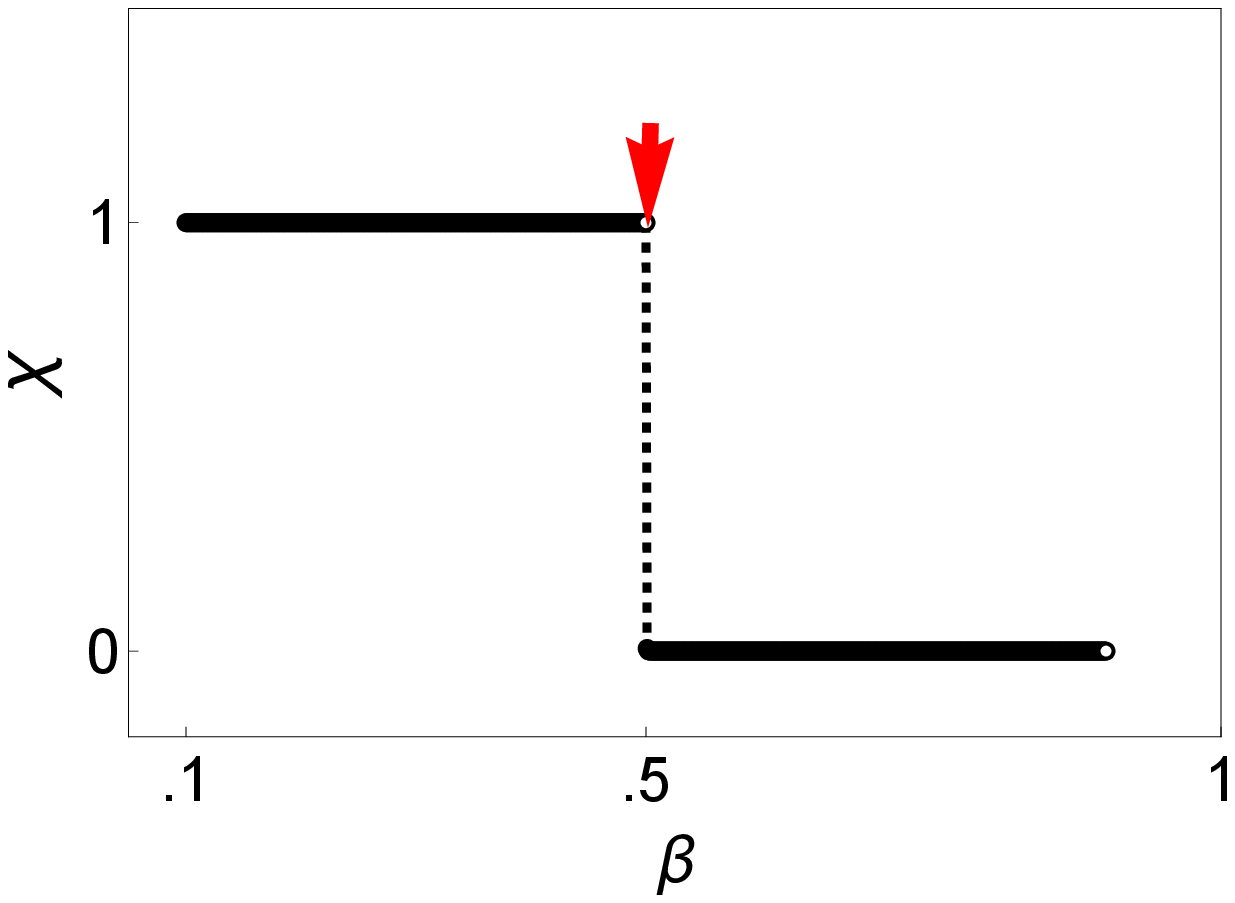} \\
  \textbf{(c)}  &   \textbf{(d)} \\
  \includegraphics[height=3cm, width=3.6cm, valign=c]{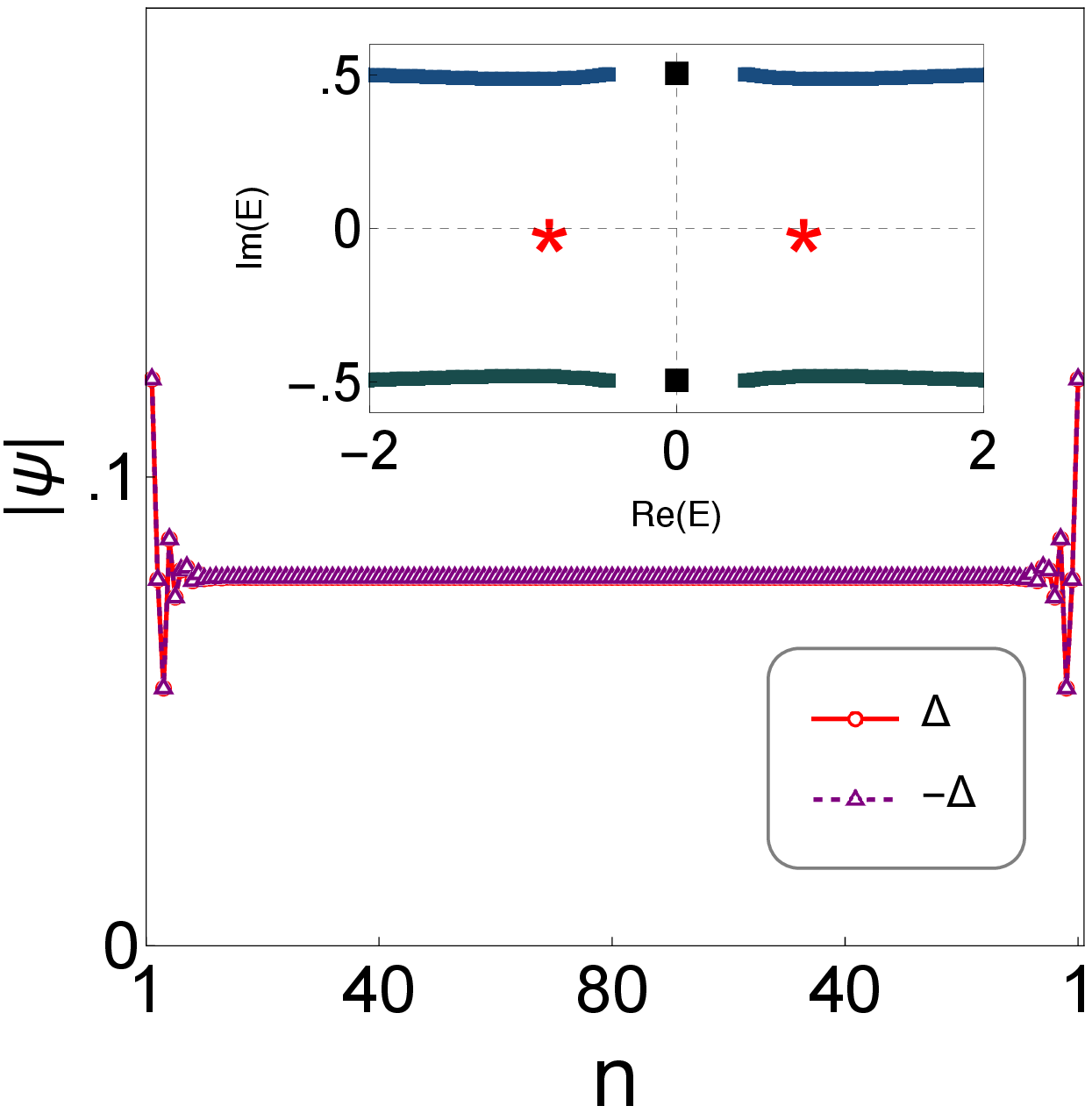}  &    \includegraphics[height=3cm, width=3.6cm, valign=c]{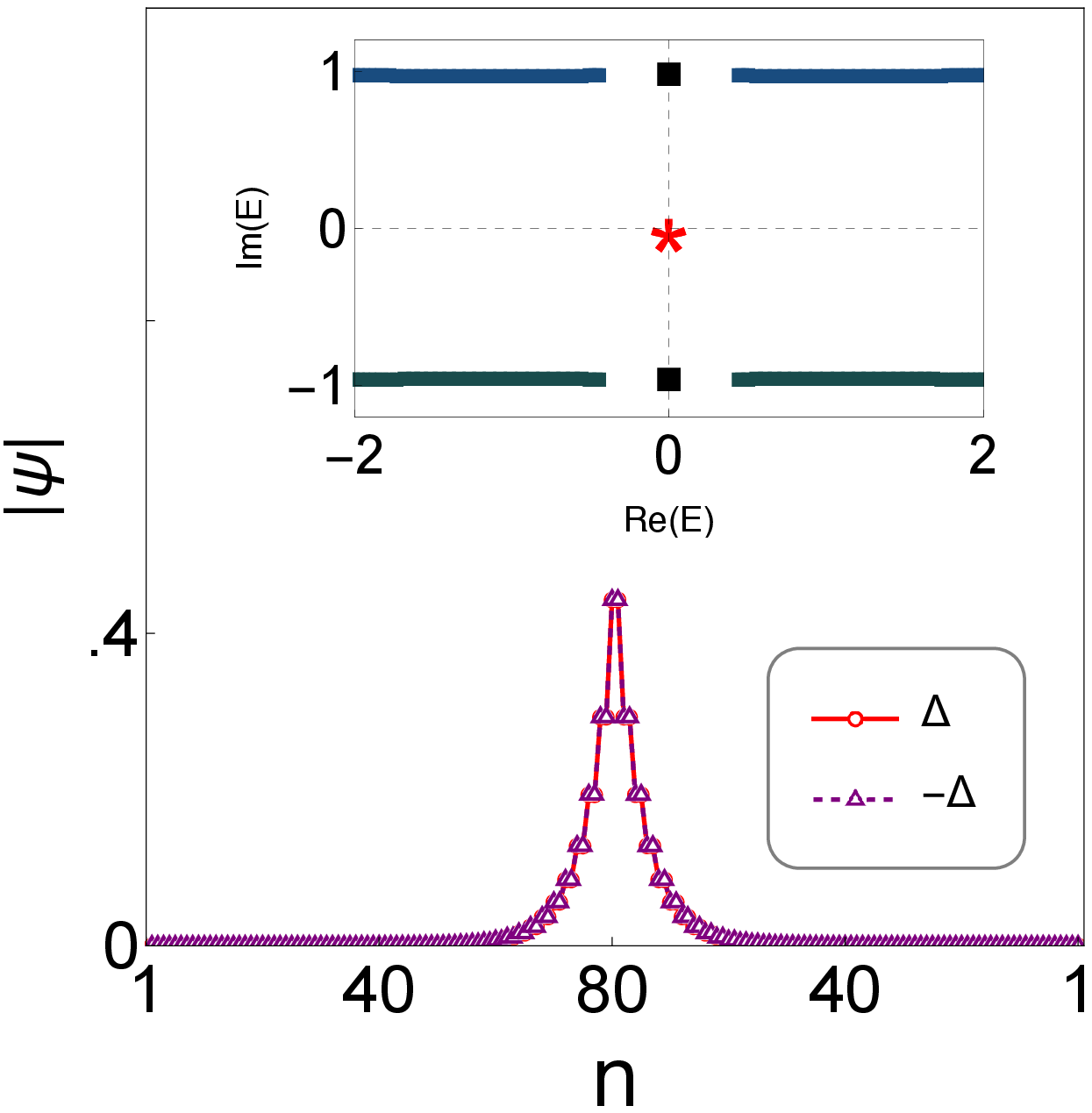} \\
  \end{tabular}
  \caption{ (a) Absolute value of the energy difference between two modes $\pm\Delta$ as a function of $\beta$ from numerical results (black circles) and analytical results (red triangles). (b) Variation of $\chi(\Delta)$ with $\beta$. (c)-(d) Spatial distribution of eigenvectors $\psi(\pm\Delta)$ and eigenvalues (inset) from (\ref{eq1}), for t$\neq$$t_{2}$. Panel (c) for $\beta$=$\gamma$ and (d) for $\beta$=1.94$\gamma$. We choose $t_{2}$=t in (a),(b) and $t_{2}$=1.3t in (c),(d). The other parameters in all panels are $\gamma$=0.5, $t_{c}$=$t_{2}$, and N=40.}
 \label{f2}
\end{figure}

The Hermiticity of a Hamiltonian, which connects a system with the physical realm, always ensures the conservation of particles and energy. However, this property can break down in open systems where interaction with the environment leads to an effective non-Hermitian description~\cite{bender2007making,bender2002complex,ashida2020non}, characterized by complex energy eigenvalues and non-orthogonality of eigenvectors~\cite{dattoli1990non,keck2003unfolding}. The introduction of non-Hermiticity in a simple lattice system can be achieved in two ways: by balancing on-site imaginary gain-loss potential~\cite{cerjan2018effects,zhu2014pt,liu2020gain,song2019breakup} or introducing asymmetric directional hopping between two sites ($t_{ab} \neq t_{ba}$)~\cite{longhi2015non,longhi2017non,song2019non,ghatak2020observation,helbig2020generalized}.\\
\indent{} With the recent introduction of non-Hermiticity to well-explored topological phases of matter in condensed matter systems~\cite{jackiw1976solitons,su1979solitons,su1980soliton,haldane1988model,kane2005quantum,hasan2010colloquium,bernevig2006quantum}, topological physics has advanced beyond the Bloch band theory \cite{bansil2016colloquium}. A fascinating phenomenon called the non-Hermitian skin effect (NHSE) \cite{okuma2020topological,lee2019anatomy,alvarez2018non,yokomizo2021scaling,zhang2021acoustic,zhang2021observation,liang2022dynamic,li2022dynamic,lee2019anatomy,yao2018non,yao2018edge,yang2020non,yokomizo2019non,deng2019non,song2019non,edvardsson2019non,longhi2019topological,kawabata2019symmetry,lee2019topological,zhu2020photonic,kunst2018biorthogonal,franca2022non,alvarez2018non} has recently been observed. It describes the wave localization of all eigenmodes toward the open boundaries~\cite{helbig2020generalized,ghatak2020observation} causing substantial difference in energy spectra between an open and a periodic chain \cite{helbig2020generalized,lee2019anatomy}. Its appearance significantly modifies the conventional Hermitian bulk-boundary correspondence (BBC)~\cite{hatsugai1993chern,essin2011bulk,rudner2013anomalous} and opens up new research avenues for non-Hermitian topological phases of matter in the form of generalized BBC~\cite{yao2018edge,yang2020non,yokomizo2019non,deng2019non,song2019non,edvardsson2019non,longhi2019topological,kawabata2019symmetry,lee2019topological,zhu2020photonic,kunst2018biorthogonal}.\\
\begin{figure*}[tb] 
\begin{center}
\begin{tabular}{c}
\begin{tabular}{c c}
  \textbf{(a)}  \\
  \includegraphics[height=\y cm, width=11 cm, valign=m]{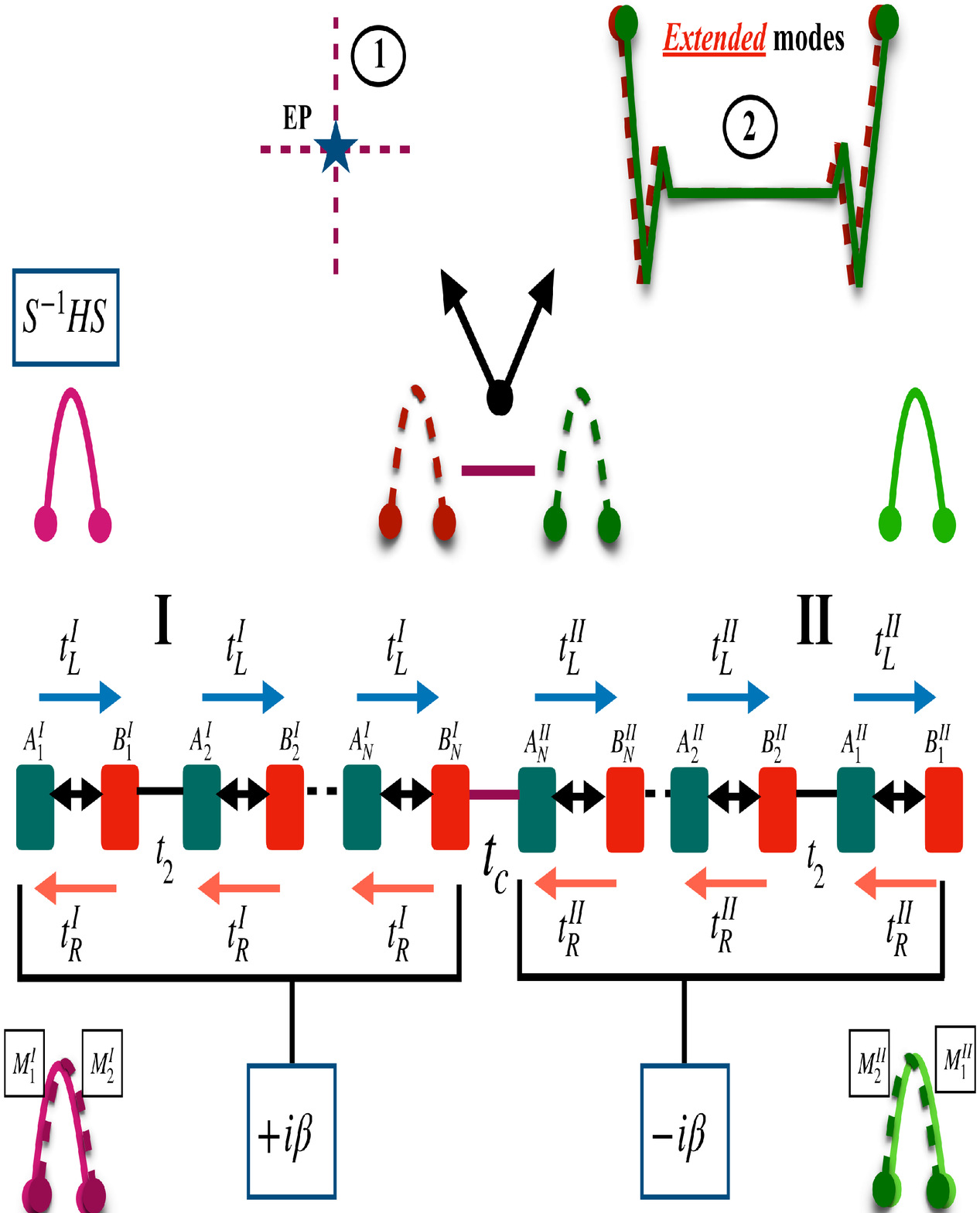} 
  & \begin{tabular}{c} 
\textbf{(b)} $t_{c}$=0 \\
\includegraphics[height=3.5cm, width=4.0cm, valign=t]{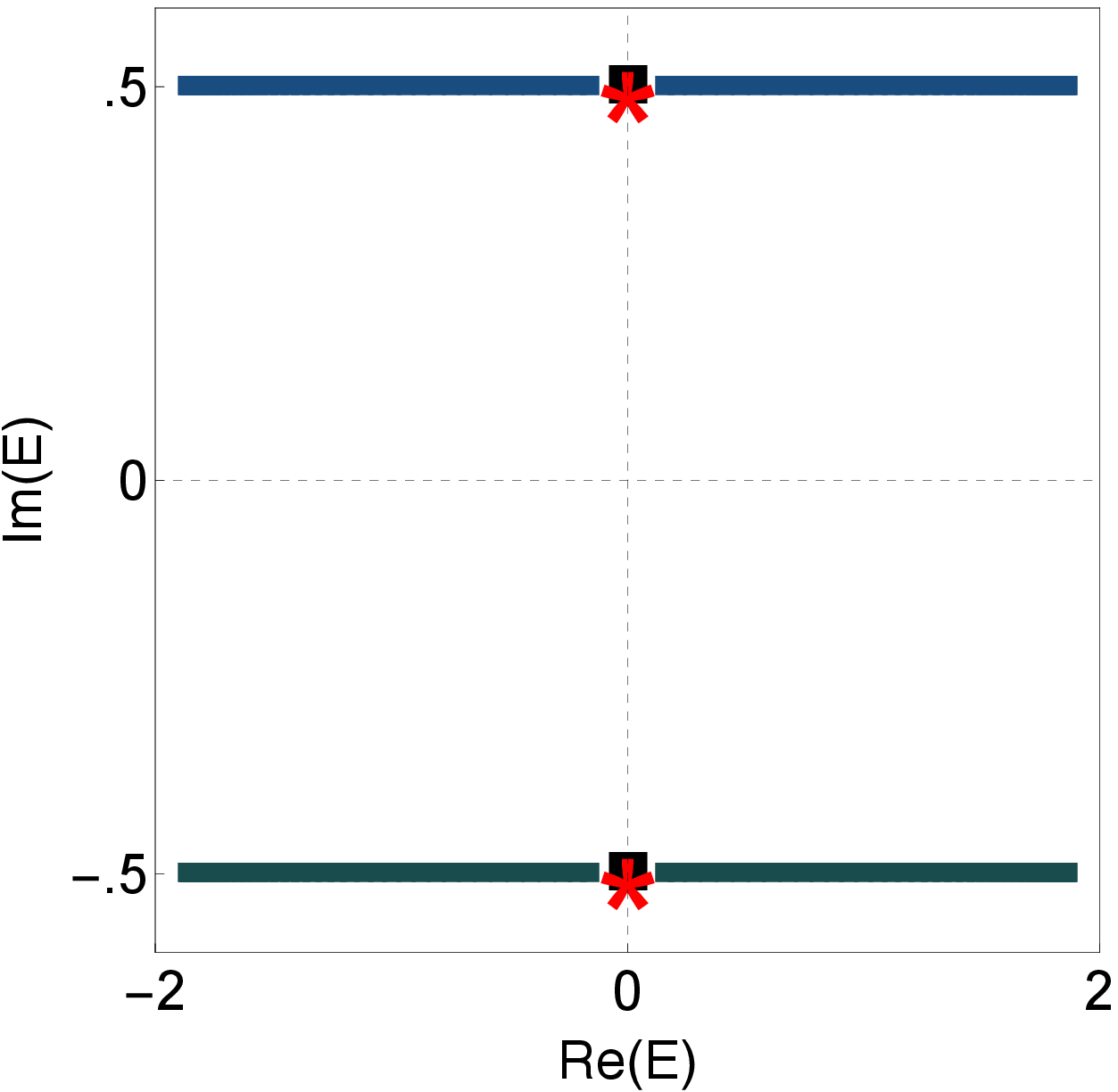}\\
\\
\textbf{(c)} $t_{c}$$\neq$0\\
\includegraphics[height=3.5cm, width=4.0cm]{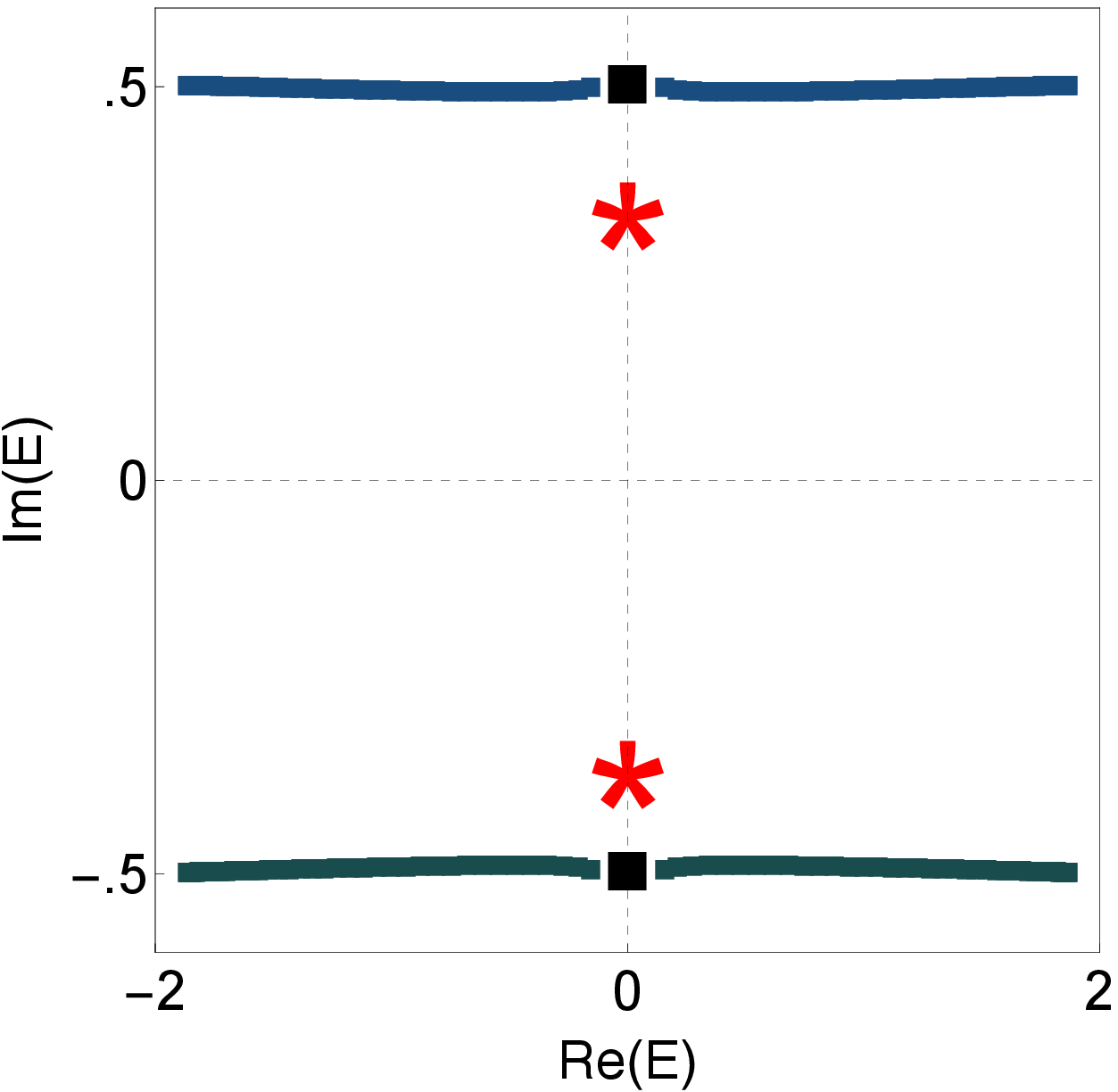}\\
\end{tabular}
\end{tabular}\\
\\
\textbf{(d)  Eigenenergy spectra} \\
\\
\begin{tabular}{c c c c}
% & \textbf{(b) Eigen energy spectra} &  &  \\
% \\
\textbf{$\beta$=0.2$\gamma$} & \textbf{$\beta$=0.6$\gamma$} & \textbf{$\beta$=$\gamma$} & \textbf{$\beta$=1.4$\gamma$} \\
\includegraphics[height=\x cm, width=4.0cm, valign=c]{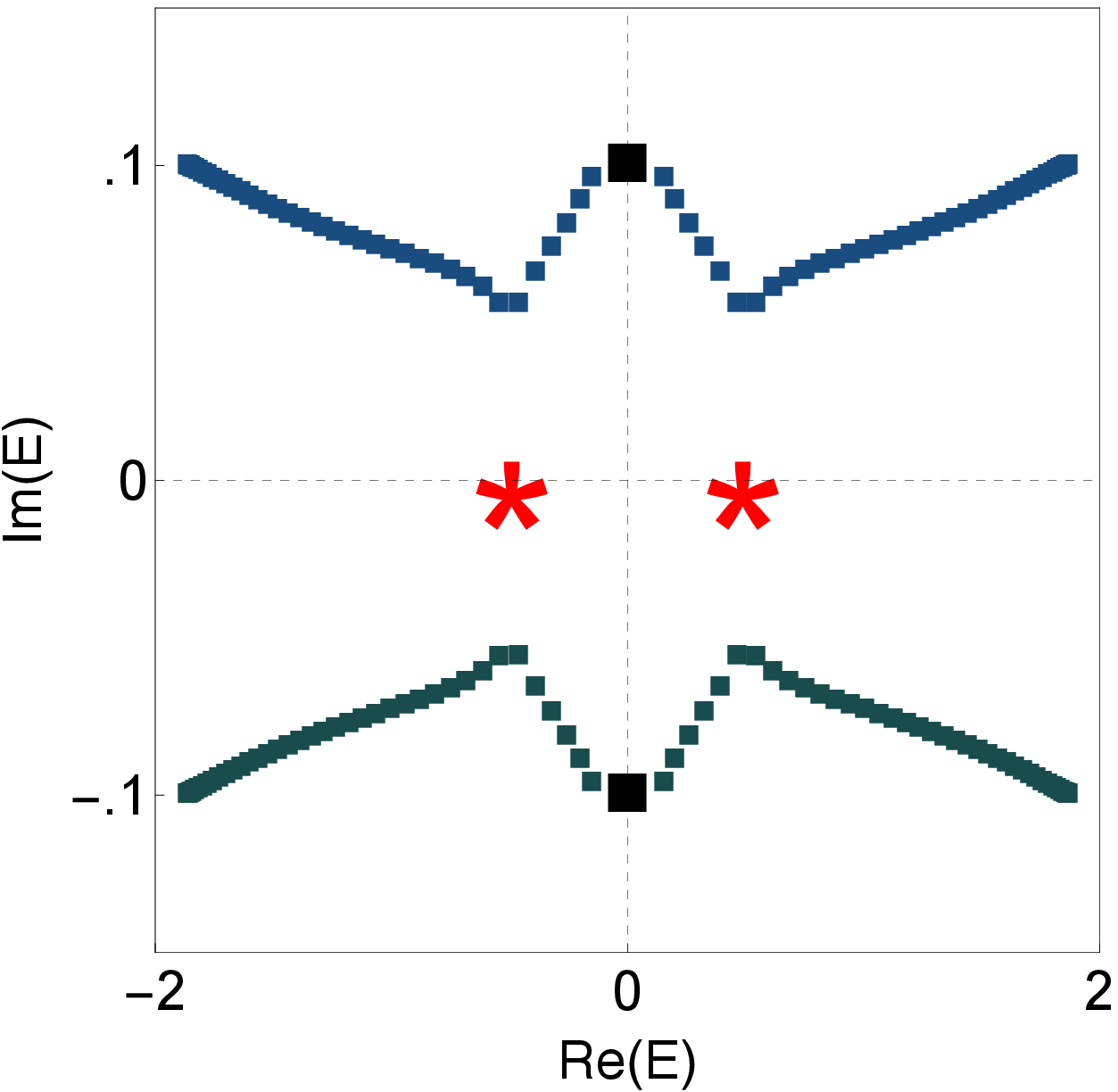} & \includegraphics[height=\x cm, width=4.0cm, valign=c]{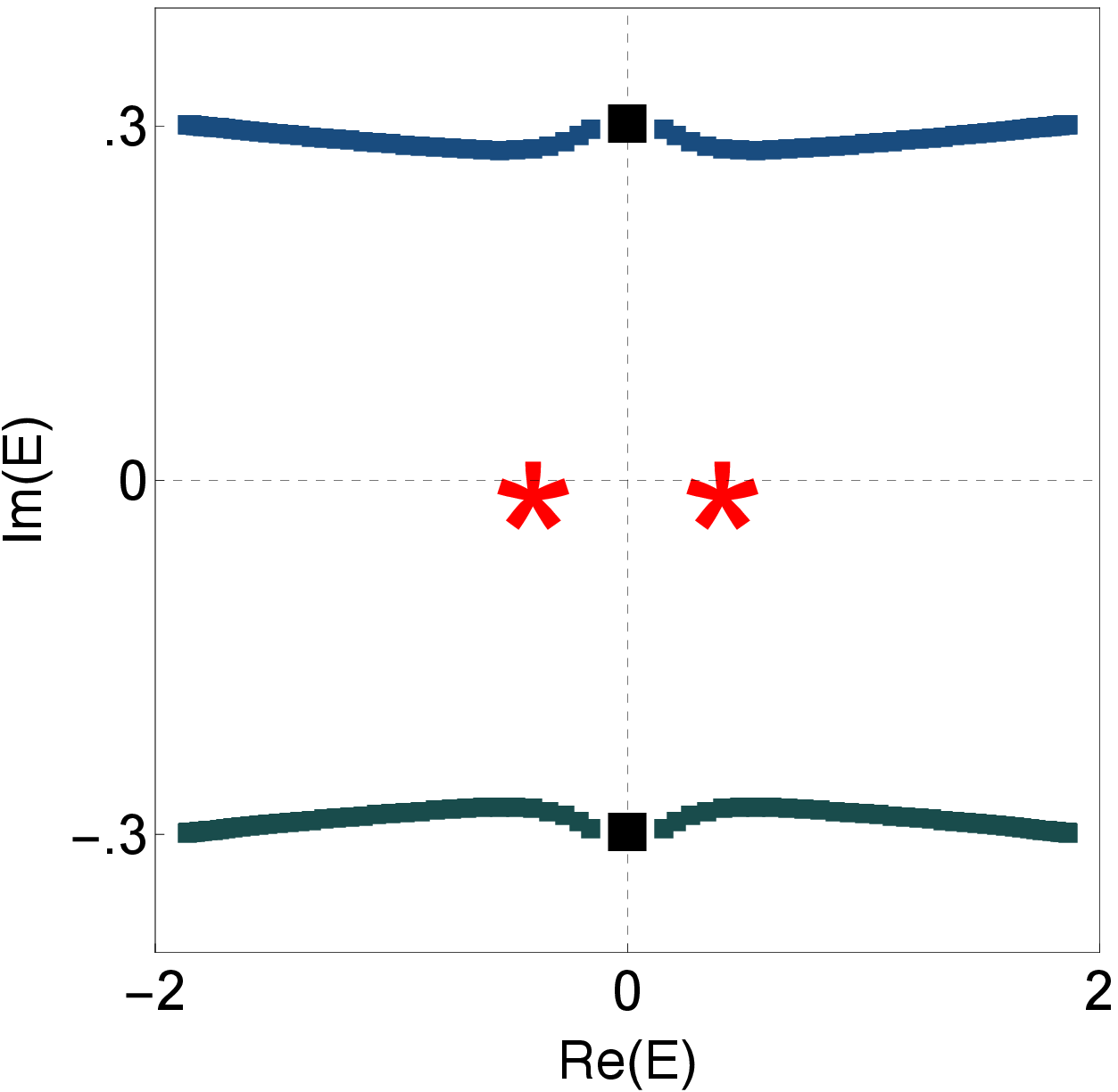} & \includegraphics[height=\x cm, width=4.0cm, valign=c]{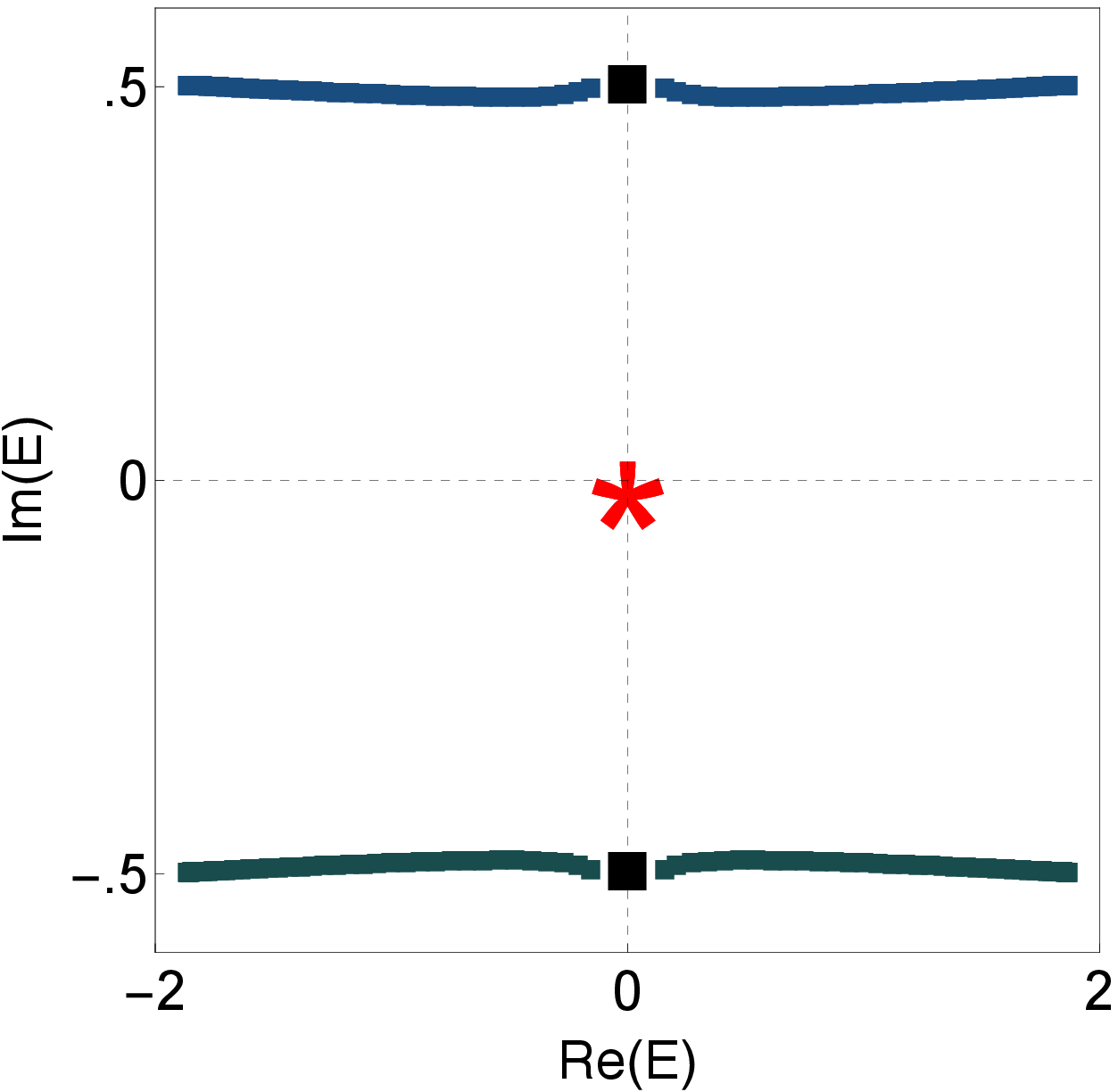} &
 \includegraphics[height=\x cm, width=4.0cm,, valign=c]{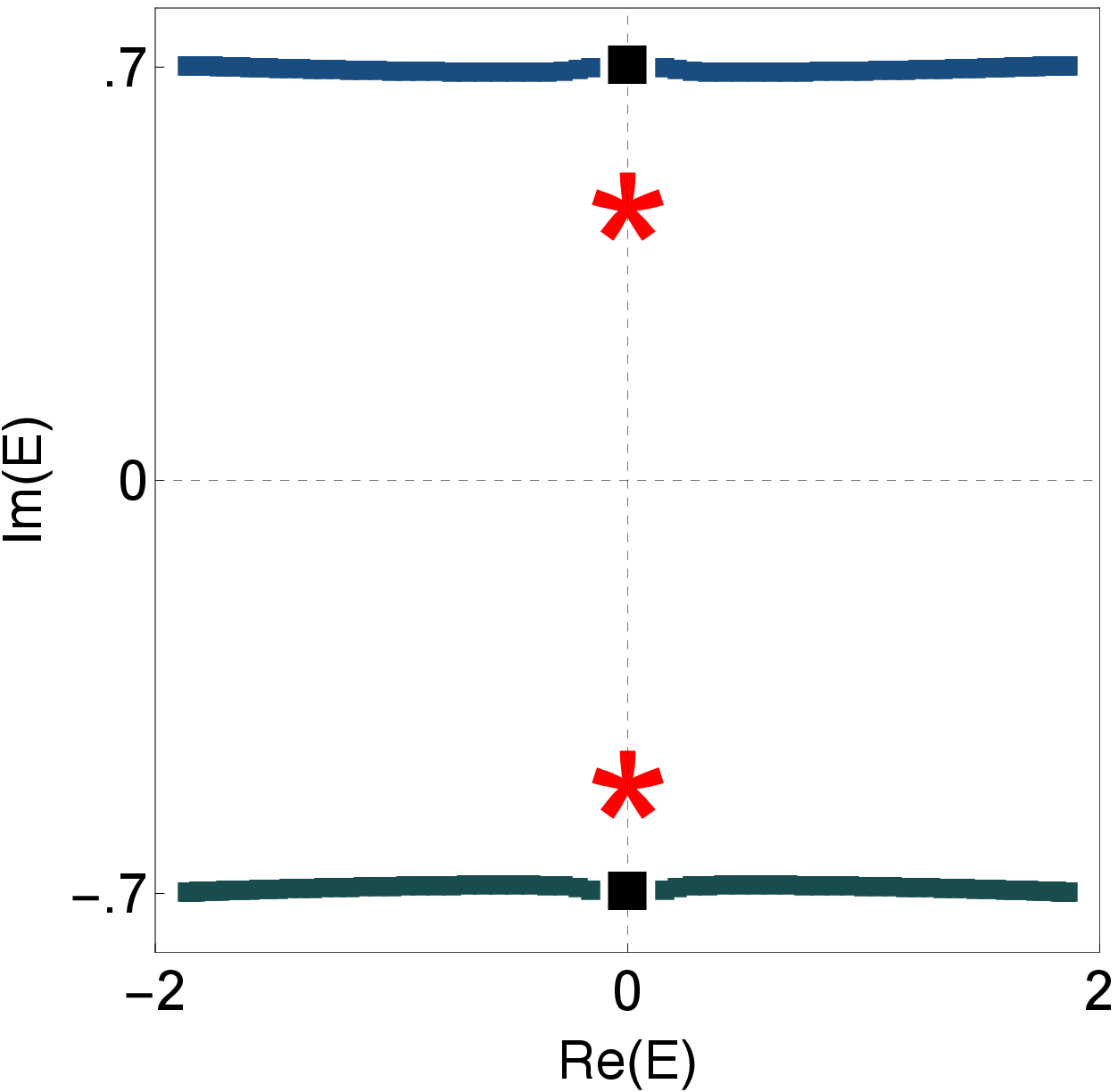}  
\end{tabular} \\
\\
  \textbf{(e) Eigenfunctions distribution}\\
\begin{tabular}{c c c c}
\textbf{} & \textbf{} & \textbf{} & \textbf{} \\
\includegraphics[height=\x cm, width=4.0cm, valign=c]{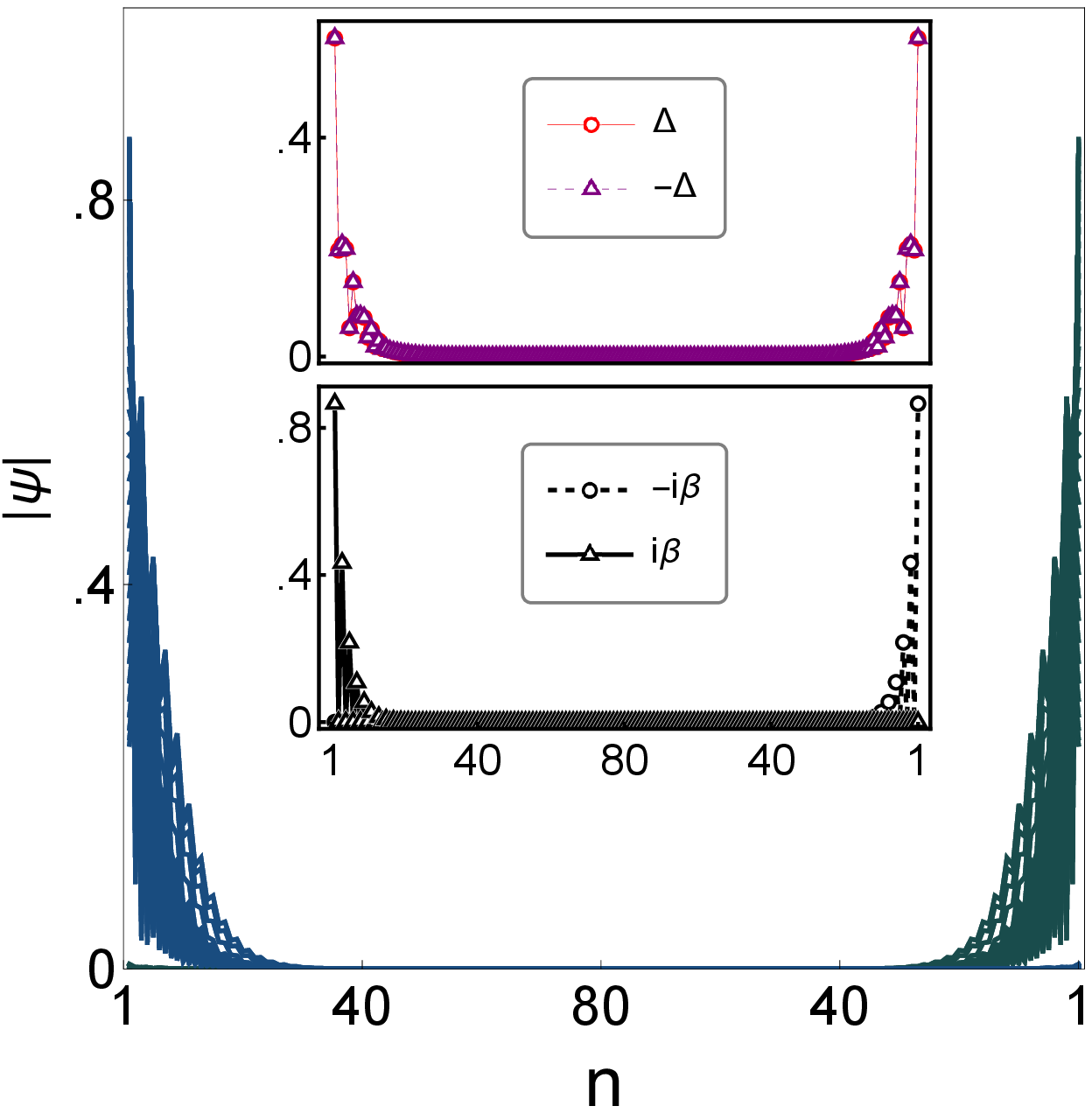} & \includegraphics[height=\x cm, width=4.0cm, valign=c]{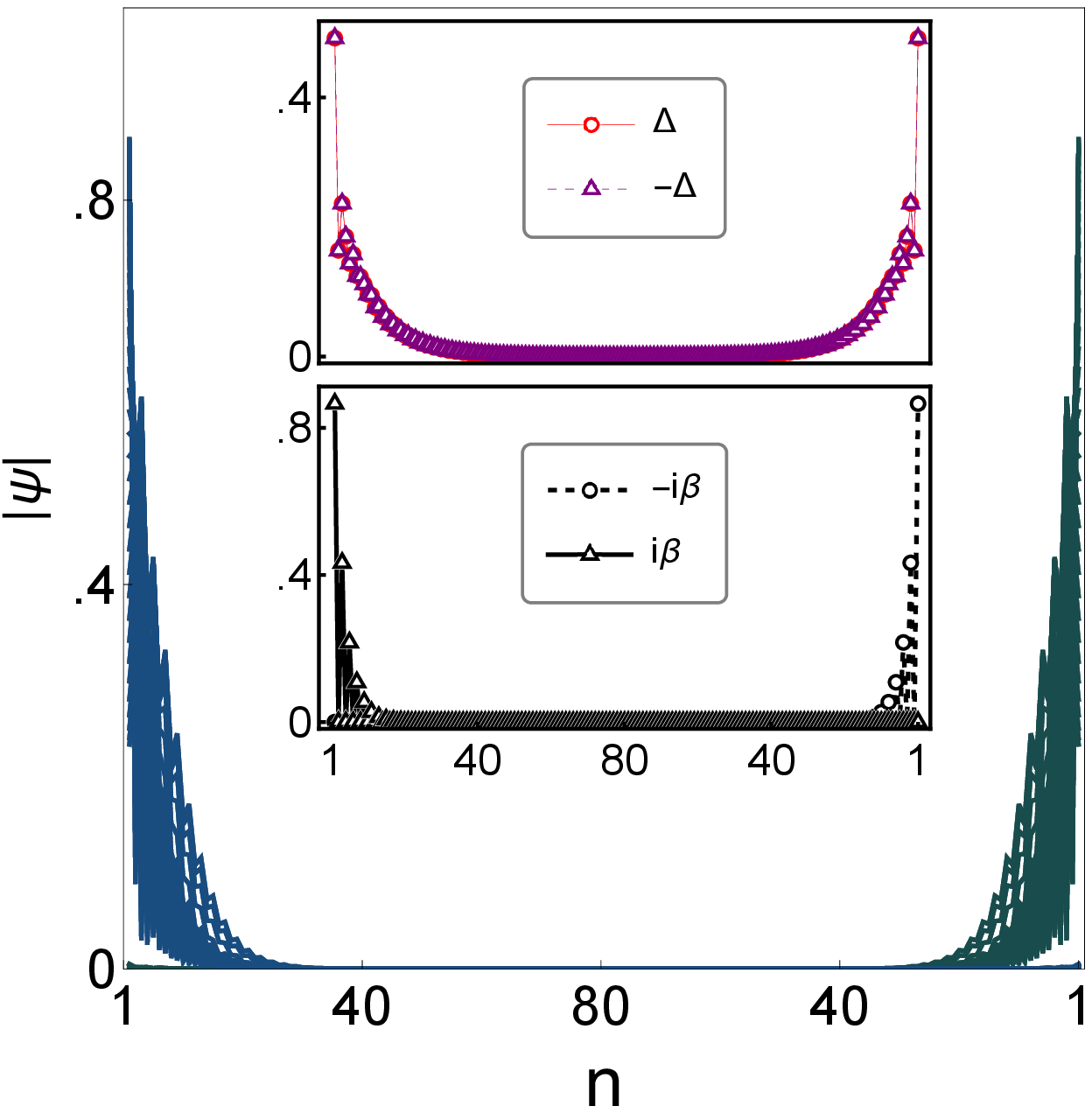} & \includegraphics[height=\x cm, width=4.0cm, valign=c]{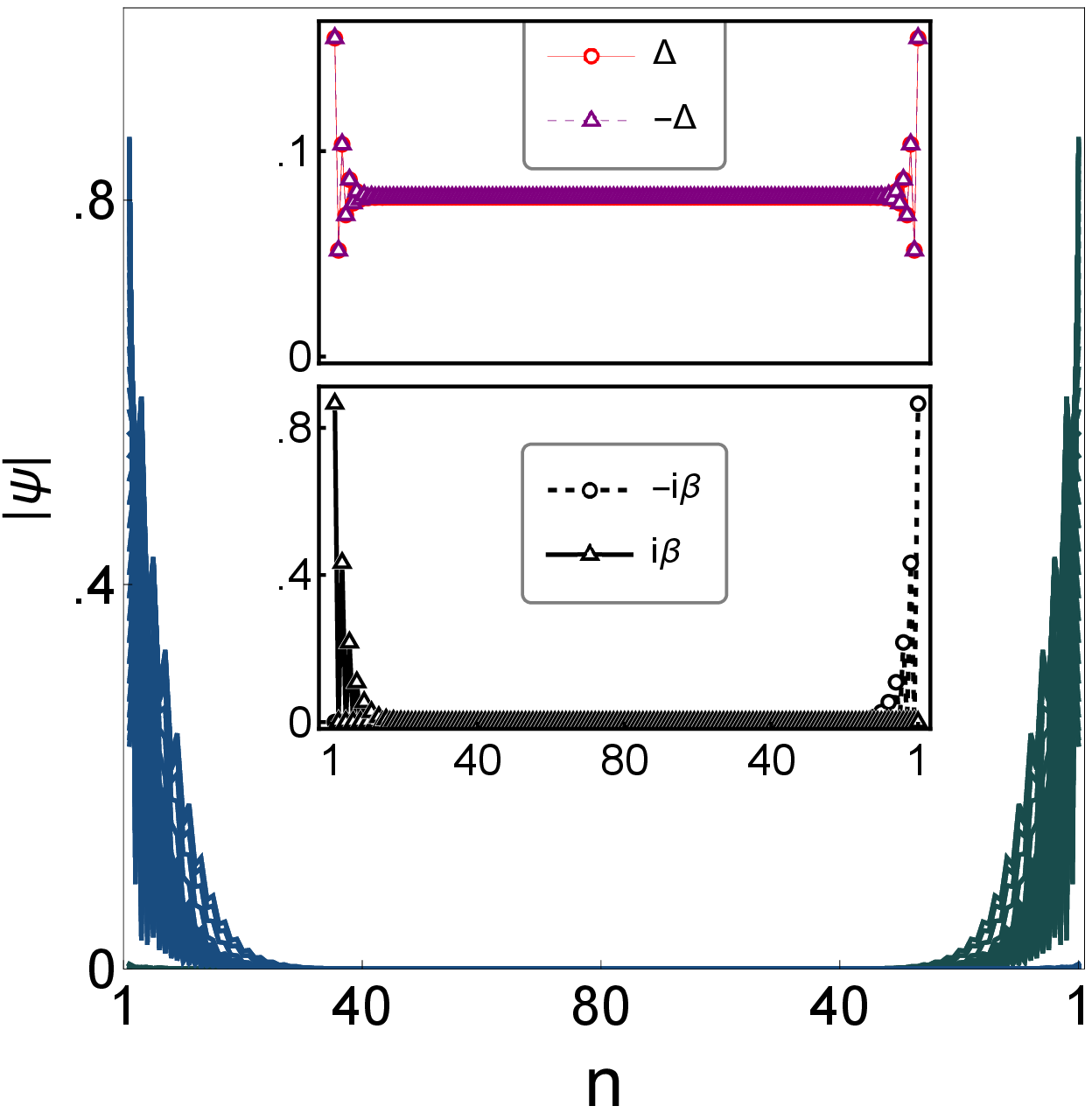} &
 \includegraphics[height=\x cm, width=4.0cm,, valign=c]{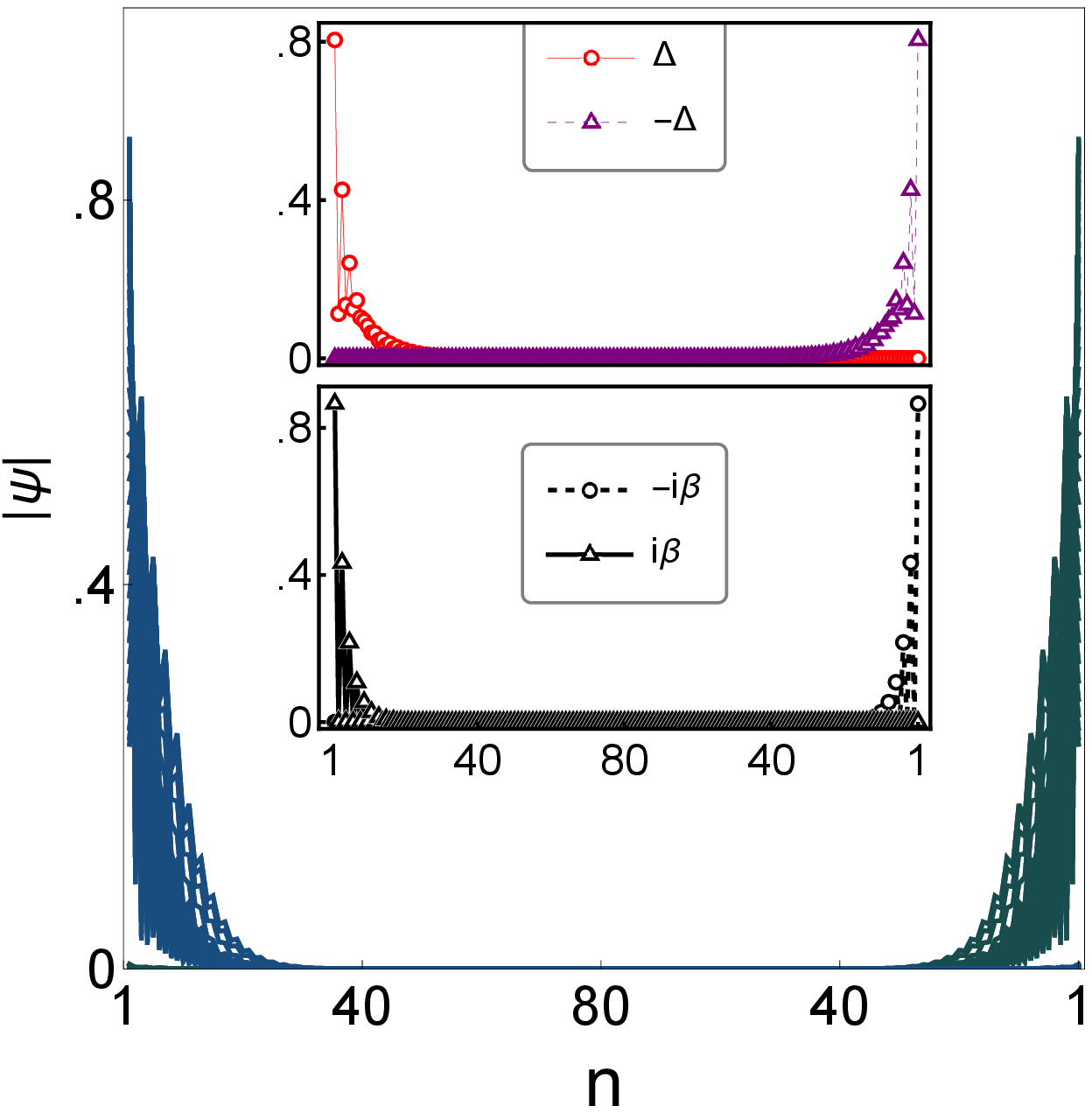}   
 \end{tabular}
 \end{tabular}
\caption{(a) An illustration of the interface between two non-Hermitian SSH chains I and II. Two chains are connected by $t_{c}$ in maroon. The two sublattices A and B are indicated by the black and red bars, respectively. Intracell directional couplings $t_{L/R}^{I/ II }$ are shown as the blue and orange lines. $t_{2}$ in black denotes the coupling between unit cells. Gain (i$\beta$) in chain I and loss (-i$\beta$) in chain II are added on both the sub lattices. (b)-(c) Evolution of the energy eigenvalues of the coupled non-Hermitian SSH Hamiltonian in (\ref{eq1}) as a function of the coupling between chains $t_{c}$, where $\beta$ is fixed at 0.5. (b) For $t_{c}$=0 and (c) for $t_{c}$=0.7t. (d) Energy eigenvalues of the coupled non-Hermitian SSH Hamiltonian of the double chain in (\ref{eq1}) with increasing value of $\beta$ for fixed $t_{c}$=$t_{2}$. (e) Spatial distribution of the eigenfunctions on the lattice sites, associated with eigenenergies in (d). In all panels, $t$=1, $t_{2}$=1, $\gamma$=0.5, and N=40.}
\label{f1}
\end{center} 
\end{figure*}
\indent{}Competition between topological localization and the NHSE can produce some nontrivial effects, such as delocalization of topological modes \cite{zhu2021delocalization} and hybrid skin-topological effects \cite{zou2021observation,lee2019hybrid,li2022gain}. In the latter case, the skin effect only affects the topological edge modes, resulting in corner localization, while the bulk modes remain unaffected. This type of unusual skin mode has recently been observed in two-dimensional systems in presence of non-reciprocal couplings \cite{zou2021observation,lee2019hybrid}, as well as implementing gain/loss in Haldane model \cite{li2022gain}. The question of whether qualitatively novel phases can be established by the simultaneous interplay of topology, gain-loss, and directional couplings will therefore be interesting to explore. \\
\indent{} In this Letter, we investigate the interplay of two distinct non-Hermitian couplings: asymmetric directional hopping and imaginary gain-loss potential on topological boundary modes. We discover that although the properties of the individual topological modes remain unaltered, their hybridization leads to two novel phenomena. First, by tuning two non-Hermitian couplings, the energies of both modes merge into an Exceptional Point (EP)~\cite{keck2003unfolding,miri2019exceptional,tang2020exceptional,kato1966perturbation,ding2022non}, which acts as a branch point singularity~\cite{dattoli1990non,keck2003unfolding} in the parameter space of non-Hermitian couplings carrying half integer topological charge. Second, we can control the distribution of eigenfunctions of the hybridized modes. Interestingly, when the strength of these couplings completely matches, NHSE localization due to asymmetric hopping is completely overcome, and the modes get extend over the bulk lattice space.  Our numerical findings are strongly supported by analytical calculations.\\
\indent{}In our model, we consider a junction between two 1D non-Hermitian Su-Schrieffer-Heeger (SSH)~\cite{guo2021exact,yao2018edge} chains with different non-reciprocal intracell coupling. The schematic of our model is shown in Fig.~\ref{f1}(a). Each chain I and II consists of two sublattices $A$ and $B$ within a unit cell. In both chains, we considered an equal number of unit cells (N). The strength of intercell coupling is represented by $t_2$, and the directional non-reciprocal intracell couplings in both chains I and II, are represented by $t_{L/R}^{I/ II }$. We choose $t_{L/R}^{I/ II }$= $t\pm{\gamma}^{I/ II }$ and $\gamma^{I/II}$ to be opposite to each other, i.e., $\gamma^{I/ II }$= $\pm{\gamma}$. $\gamma$ denotes non-reciprocity strength. We also consider the onsite gain and loss on all sublattices in terms of imaginary onsite potential $\pm i\beta_{}$ in chain I and II, respectively. The coupling $t_{c}$ connects the two chains at one end. We start with the 1D Hamiltonian of the tight-binding model of the coupled chain in real space, which has the following matrix form:
\begin{equation} \label{eq1}
H_{s}= 
\begin{bmatrix}
  H_{I} &
  t_{c}\\ 
  t_{c} & H_{II}
\end{bmatrix},
\end{equation} 
where $H_{I}$ and $H_{ II }$ denote Hamiltonian for the chains I and II, respectively . The Bloch Hamiltonian for the uncoupled chains reads
\begin{equation} \label{eq2}
H_{I/II}^{k}=\alpha_{x}\sigma_{x} + (\alpha_{y}\mp \gamma) \sigma_{y} \pm i\beta \sigma_{0},
\end{equation}
where $\alpha_{x}$=$t$+$t_{2}\cos(k)$, $\alpha_{y}$=$t_{2}\sin(k)$, $\sigma_{x,y}$ are the Pauli matrices and $\sigma_{0}$ is the Identity matrix. We present a detailed discussion of the characteristics of a single chain with directional couplings in the presence of gain, loss in the supplementary material~\cite{SM}(\ref{SA}).\\
\indent{}In the main text we study the scenario $t_c$$\neq$0, i.e., the two chains are coupled at one end by the coupling $t_{c}$. We set the strength of the $t_2$ coupling as $t_{2}$$ > $$\pm\sqrt{t^{2}-\gamma^2}$, so that both chains remain individually in the topological regime and two topological modes exist in each chain at the open boundaries. When $t_{c}$=0, as shown in Fig.~\ref{f1}(b), the energy bands in blue and green  obtained from the numerical diagonalization of (\ref{eq1}) correspond to chain I and chain II. Two edge modes for each chain exists inside the bulk gap are represented by black bars and red stars, respectively. Now, gradually increasing the strength of $t_{c}$, the modes in black remain at ($\pm i\beta$), but the modes in red stars shift toward the origin, as shown in Fig.~\ref{f1}(c). These modes in red are labeled $\pm \Delta$.\\
\indent{} This splitting of the modes from their initial energy when coupling $t_{c}$ is introduced is to be understood as follows: In the uncoupled chain $H_{I/II}$, NHSE forces the topological zero modes to be localized at one end ~\cite{SM}(\ref{SA}) due to the nonreciprocal couplings, as shown in the schematic diagram of Fig.~\ref{f1}(a). The topological modes in chain I are denoted as $M_{1,2}^I$ and in II as $M_{1,2}^{II}$. Now one can perform a similarity transformation as $S^{-1}HS$ on the Hamiltonians $H_{I/ II }$ and make them Hermitian, leaving the eigenvalues unchanged and the eigenvectors shifting their position ~\cite{SM}(\ref{SA}) to both open ends. The eigenfunctions $\psi_{1,2}$ of the zero modes in each of the chains satisfy the following boundary conditions~\cite{chen2020elementary}: $\psi_{1B,NA}^{I}$=0 in chain I and $\psi_{1A, NB }^{ II }$=0 in chain II. If we now introduce $t_{c}$, the boundary conditions $\psi_{1B}^{I}$=0 in chain I and $\psi_{1A}^{II}$=0 in chain II remain unchanged, and two modes at zero energy at the open ends with zero weight $\psi_{n(B/A)}^{I/II}$=0 remain, i.e., these modes weight on sublattice A in chain I and on sublattice B in chain II, respectively. The introduction of gain and loss shifts their energy as ($\pm i\beta$). +i$\beta$ remain at the left end of chain I, which satisfies $\psi_{1B}^{I}$=0, and -i$\beta$ remain at the right end of chain II, which satisfies $\psi_{1A}^{II}$=0.\\
\indent{} However, as the initial boundary conditions change as $\psi_{NA}^{I}$$\neq$0 and $\psi_{NB}^{II}$ $\neq$0, upon introducing $t_{c}$ $\neq$0 the other two zero modes hybridized into finite-energy states. The introduction of gain and loss allows the gain on the left side and loss on the right side for both hybridized modes, and the evolution of their energy and wave functions are discussed below.\\
\indent{}In Fig.~\ref{f1}(d) we demonstrate the eigenenergy spectrum at fixed $t_{c}$=$t_{2}$ for the range of strength $\beta$, from small $\beta$ = 0.2$\gamma$ to large $\beta$ = 1.4$\gamma$, through the point $\beta$ = $\gamma$. Two
bands are shown in blue in the upper half of the complex plane and the other two in green in the lower half. As expected two modes in black with the energy ($\pm i \beta$) always exist in the gap between the upper and lower bulk bands.
The other modes $\pm \Delta$ initially lie on the real energy axis (panel. $\beta$=0.2$\gamma$). The energy difference $|2\Delta|$ steadily decreases with increasing strength of $\beta$ (panel. $\beta$=0.6$\gamma$) until the critical point where $\beta$=$\beta_{c}$, the energy difference between these modes becomes zero. At $t$=$t_{2}$, $\beta_{c}$ is equal to the strength of the non-reciprocity $\gamma$. For larger values away from the critical point ($\beta$$>$$\beta_{c}$), the energies becomes imaginary, as shown in the panel for ($\beta$=1.4$\gamma$). Thus, as a function of $\beta$, there is a shift of the energy from the real to the imaginary axis, with a gap closing at $\beta_{c}$. This transition is particularly distinctive, and the crossing point is referred the Exceptional point (EP) \cite{keck2003unfolding}.\\
\indent{} The corresponding eigenfunction distribution is illustrated in Fig.~\ref{f1}(e).  All the bulk states with energy in the positive imaginary axis in blue are accumulated at the left end of I, while bulk states with energy in the negative imaginary axis are accumulated at the right end of II, as shown in green, owing to NHSE ~\cite{SM}. Also as expected, states with energies $\pm i \beta$ are localized in black at the left and right ends, unchanged by $\beta$, showing typical topological edge state behavior~\cite{SM}.\\
\indent{} A very interesting characteristics at EP can also be observed in the eigenfunction distribution of the modes $\psi(\pm\Delta)$ shown in the inset of Fig.~\ref{f1}(e). These modes are initially (panel. $\beta$=0.2$\gamma$) uniformly confined to the two open ends of chains I and II. Now, as the energy difference $|2\Delta|$ gradually decreases with the increase of $\beta$, it correlates with a steady decrease of the localization length (panel. $\beta$=0.6$\gamma$) until they are completely delocalized at EP, $\beta$=$\gamma$. A further increase (panel. $\beta$=1.4$\gamma$) away from EP transfers the $+\Delta$ mode to chain I and -$\Delta$ to II. Thus, the complete transfer of the eigenfunction distribution from both chains to the single chain I for the +$\Delta$ mode and to the chain II for the -$\Delta$ mode via the EP is found to be a distinctive signature of EP, and is one
of the important findings of this work.\\
\indent{}To better understand the interplay between gain-loss strength ($\beta$), non-reciprocity ($\gamma$), and topology ($t,t_{2}$), we show in the supplementary material~\cite{SM}(\ref{SB}) an analytic framework for determining the eigenenergies and eigenfunctions by solving (\ref{eq1}). In particular, assuming that the system size is very large, i.e., N $\gg$ 1, we obtain the expressions for the eigenenergies of the modes $\pm \Delta$ as,
\begin{equation} \label{eq3}
\Delta =  i \sqrt{t^2-t_{2}^2+\beta^2-\gamma^2}.
\end{equation}
From (\ref{eq3}) it can be seen that for the specific parameter values $t_2$=t and $\beta$=$\gamma$ the square root vanishes, indicating the formation of EP. In Fig.~\ref{f2}(a), the absolute value of the energy difference $|\delta E|$=$|2\Delta|$ between the modes ($\pm\Delta$) is shown as a function of $\beta$ numerically (black circles) and analytically (red triangle). Our analytical formula (\ref{eq3}) is confirmed by the numerical results.\\
\indent{}The exact expressions for the components of the right eigenfunction~\cite{SM} in the I and II chains of the mode +$\Delta$ are as follows:
    \begin{align}\label{eq4}
    \begin{cases}
\psi_{nA}^{I}(\Delta)= \Bigg[ \frac{\mu_{1}^{}}{\mu_{2}^{}}\left(\frac{-\mu_{2}^{}}{t_{L}^{I}}e^{i \theta}\right)^n+ \left(\frac{-\mu_{1}^{}}{t_{L}^{I}}e^{-i\theta}\right)^n\Bigg]\Phi_{A1}^{I},   \\
   \psi_{nB}^{I}(\Delta)=\frac{t_{R}^{I}}{i \mu_{2}^{}}\Bigg[ \left(\frac{-\mu_{2}^{}}{t_{L}^{I}}e^{i \theta}\right)^n - \left(\frac{-\mu_{1}^{}}{t_{L}^{I}}e^{-i\theta}\right)^n\Bigg]\Phi_{A1}^{I}, \\
    \psi_{nB}^{II}(\Delta)=\Bigg[ \frac{\mu_{1}^{}}{\mu_{2}^{}}\left(\frac{-\mu_{2}^{}}{t_{R}^{II}}e^{-i \theta}\right)^n+ \left(\frac{-\mu_{1}^{}}{t_{R}^{II}}e^{i\theta}\right)^n\Bigg]\Phi_{B1}^{II},\\
    \psi_{nA}^{II}(\Delta)=\frac{t_{L}^{II}}{i \mu_{2}^{} }\Bigg[ \left(\frac{-\mu_{2}^{}}{t_{R}^{II}}e^{-i \theta}\right)^n - \left(\frac{-\mu_{1}^{}}{t_{R}^{II}}e^{i\theta}\right)^n\Bigg]\Phi_{B1}^{II} ,
    \end{cases}
    \end{align}
where \textit{n} is the unit cell index,  
 $\tan{\theta}$=$\Delta$/$\sqrt{t^2 + \beta^2 - \gamma^2}$,  $\mu_{1/2}^{}$=$\sqrt{t^2+\beta^2-\gamma^2}\mp\beta$.\\
\indent{} The signature of EP in the real space distribution of the eigenfunctions in inset of Fig.~\ref{f1}(e), can be revealed by plotting the behavior of $\chi$=$|\phi_{B1}^{ II }/\phi_{A1}^{I}|$, which determines the projection of any eigenfunction onto the chain II (details in the supplementary material ~\cite{SM}, (\ref{sq13})). We plot $\chi(\Delta)$ as a function of $\beta$ in Fig.~\ref{f2}(b) and observe that it has a constant value of `1' throughout the range $\beta$$\leq$$\gamma$, indicating that this mode is equally distributed in both chains. In contrast, there is a sharp transition to `0' for $\beta$$ > $$\gamma$, indicating that this mode was distributed only in the I chain. Similarly, $\chi(-\Delta)$ as a function of $\beta$ shows that the other mode with energy -$\Delta$ remains concentrated in the II chain for $\beta$$>$$\gamma$. \\
\indent{}The delocalization of the modes $\psi(\Delta)$ at $\beta$=$\gamma$ can be understood from (\ref{eq4}). In the region  $\beta$$\leq$$\gamma$ , $\theta$ is real and $|e^{\pm i \theta}|$=1,  $|\mu_{1}/t_{L}^{I}|$$\ll$1. Hence, the localization length is solely dependent on the factor $|\mu_{2}/t_{L}^{I}|$. As the value of $\beta$ steadily grows, $\mu_{2}$ simultaneously increases and leads to gradual delocalization of the mode $+\Delta$. Now, at $\beta$=$\gamma$, $\mu_{2}$=$t_{L}^{I}$ and  $\mu_{1}$=$t_{R}^{I}$. Therefore far from the left end in chain I i.e., n$\gg$1, $|\psi_{n(A,B)}^{I}(\Delta)|$ converges to $|t_{R}^{I}/t_{L}^{I}|$. Similarly, we can obtain the convergence of $|\psi_{n(A,B)}^{II}(\Delta)|$
to $|t_{L}^{II}/t_{R}^{II}|$ in chain II. This indicates that $\psi(\Delta)$ becomes delocalized at $\beta$=$\gamma$, i.e., overcomes the NHSE. The degree of delocalization reads,
\begin{equation}
 |\psi_{n(A,B)}(\Delta)|=|(t-\gamma)/(t+\gamma)|,
\end{equation}
i.e., delocalization is constant and depends only on the relative strength of the directional couplings.  
Finally, in the region  $\beta$$>$$\gamma$,  $\theta$ turns imaginary, we have $|e^{i\theta}|<$1, $|e^{-i \theta}|>$1 and $|e^{-i \theta}(\mu_{1}/t_{L}^{I})|<$1. Hence, in the limit n $\gg$ 1, we obtain $|\psi_{n(A,B)}^{I}(\Delta)|$ $\to$ 0, demonstrating the transition from total de-localization to a localized state in the open left end. The evolution of the mode $\psi$(-$\Delta$) can be verified in similar way.\\
\indent{}Here we want to emphasize that the emergence of EP and delocalization of the modes $\psi(\pm\Delta)$ are two distinctive features of our model. t=$t_{2}$ is the special case where these two features merge. To demonstrate this we choose t$\neq$$t_{2}$ and plot eigenvalues and $\Psi$($\pm\Delta$) in Fig.~\ref{f2}(c)-(d). Fig.~\ref{f2}(c) shows that $|\psi(\pm\Delta)|$ gets delocalized at $\beta$=$\gamma$ where 2$|\Delta|$$\neq$0 which signifies these modes are away from EP and (d) at EP these modes are confined at the junction.\\
\indent{}We characterize the discrete modes $\bigl\{\pm i \beta,\pm\Delta \bigl\}$ by biorthogonal polarization \textit{P}~\cite{kunst2018biorthogonal} given as follows,
\begin{equation}\label{eq5}
P=\lim_{N\to \inf} \biggl \langle \Psi_{L} \bigg| \frac{\sum_{n=1}^N (N-n)\Pi_{n}}{N}  \bigg| \Psi_{R} \biggr \rangle
\end{equation}
where $\Pi_{n}$= $\sum_{m}\Bigl(c_{n,m}^{I\dagger}\ket{0}\bra{0}c_{n,m}^{I} + c_{n,m}^{II\dagger}\ket{0}\bra{0}c_{n,m}^{II}\Bigr) $ is the projection operator onto unit cell \textit{n}, and \textit{m} is the sub-lattice index. The right eigen state $\ket{\Psi_{R}}$ and left eigen state $\ket{\Psi_{L}}$ of the modes can be obtained exactly as
\begin{align} \label{eq6} %\tag{S12}
 \begin{cases}
   \ket{\Psi_{R}}= \mathrm{N_{R}}\sum_{n=1}^{N}\sum_{m}^{}\Bigl(\psi_{n,m}^{I,R}{c_{n,m}^{I\dagger}}^{}+\psi_{n,m}^{II,R}{c_{n,m}^{II\dagger}}^{}\Bigr) \ket{0} \\
\ket{\Psi_{L}}=\mathrm{N_{L}}\sum_{n=1}^{N}\sum_{m}^{}\Bigl(\psi_{n,m}^{I,L}{c_{n,m}^{I\dagger}}^{}+\psi_{n,m}^{II,L}{c_{n,m}^{II\dagger}}^{}\Bigr) \ket{0},
   \end{cases}
\end{align}
where $\mathrm{N_{R,L}}$ is the normalization factor and $\psi^{R}$ and $\psi^{L}$ denote the left and right eigenfunctions respectively, as obtained from (\ref{eq1}). Away from EP the eigenfunctions with eigenenergies $\bigl\{\epsilon_{i},\epsilon_{j}\bigl\}$  satisfies the following biorthogonal normalization $\braket{\psi^{L}_{i}|\psi^{R}_{j}}$=$\delta_{ij}$.\\ 
\indent{}In accordance with the biorthogonal normalization requirement, \textit{P} is quantized for every topological boundary state regardless of the details, although the biorthogonal density $\Pi_{n}$ is typically complex valued~\cite{kunst2018biorthogonal}. Fig.~\ref{f3} demonstrates \textit{P} as function of $\beta$. We observe that \textit{P} always quantizes to 1, for both the localized modes $\pm i \beta$, which is in agreement with the signature of topological boundary modes Ref.\cite{kunst2018biorthogonal}. Away from the EP the polarization \textit{P} for the other modes $\pm \Delta$ coins at 0 as they form through the hybridization and deviate from the topological localization. At the EP due to the coalescence of the eigenstates, normalization fails, i.e., $\mathrm{N_{R}}\mathrm{N_{L}}$$\to$ $\infty$ and \textit{P} diverges.\\
\begin{figure}[tb]
\begin{tabular}{c}
  \includegraphics[height=4 cm, width=6.0cm, valign=c]{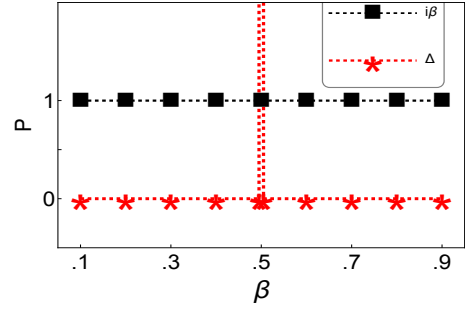} 
  \end{tabular}
\caption{Biorthogonal polarization P as a function of $\beta$ for the $\bigl\{\pm i \beta,\pm\Delta \bigl\}$ modes shown in red and black. We set $t$=1, $t_{2}$=1, $\gamma$=0.5, and N=40.}
 \label{f3}
\end{figure}
\indent{}Finally, we characterize the topology only associated with the complex energy energy dispersions in (\ref{eq3}) around the EP  by calculating the topological index vorticity $\nu$~\cite{shen2018topological}. In the parameter space of $\beta$ we defined $\nu$ as
\begin{equation}\label{eq13}
   \nu=\frac{1}{2\pi}\oint\, \partial_{\beta}\arg\biggl(\Delta_{+}(\beta)-\Delta_{-}(\beta)\biggr)\, d\beta
\end{equation}
where the integral is along a closed path in the space of $\beta$ around an a EP. We consider $t_{2}$=t at which the EP appears at $\beta$=$\pm\gamma$. To obtain a closed loop around $\gamma$ we choose a circle around $\gamma$ with radious $r$ as $\beta$=$\gamma$+r$e^{i\phi}$ in complex plane. With the following substitution we have energy difference $\delta$E=$2ie^{i\phi/2}\sqrt{2r\gamma+r^2 e^{i\phi}}$ and $\arg(\delta E)$=($\frac{\phi+\arctan(r/2\gamma)+\pi}{2}$).  Now for a single cycle of 2$\pi$ encircling the EP, one got $\Delta_{\pm}$ $\xrightarrow[]{\phi:0 \rightarrow 2\pi}$ $\Delta_{\mp}$, i.e. the two levels swap with each other in the complex plane owing to the square root singularity, and the vorticity takes the expected half-integer value.\\
\indent{} To conclude, in this letter we study the interplay of directional coupling and onsite gain, loss strength in a coupled chain network of two non-Hermitian SSH chains with opposite couplings in OBC. The un-coupled chains each have two topological modes at their boundaries. We show that fusing of two modes, one from each of the different chains, hybridize to finite energy and can coalesce to an EP depending on the strength of gain, loss and non-reciprocity. We systematically show that the bulk, and the `hybridized' modes exhibit two very different localization behaviors at the boundaries. The strength of the gain/ loss controls localization length of these modes, and can even causing them to be fully extended overcoming the NHSE.
%when the strength of the gain/loss and the directionality match. 
Analytically, we show that for a large system, the degree of delocalization becomes constant and depends only on the relative strength of the directional couplings. All discrete energy modes located within the bulk bands are characterized by biorthogonal polarization, which shows a clear signature of deviation from topological localization after hybridization. \\
\indent{} In terms of practical implementation, our model can be realized by employing topo-electric circuits~\cite{lee2018topolectrical,helbig2020generalized, hofmann2019chiral}, where non-local voltage response and impedance measurement can be utilized to detect EP and extended modes.  In most of the experiments reported, EPs are formed in momentum space by the splitting of a Dirac point into a pair of EPs~\cite{su2021direct,zhou2018observation,richter2019voigt,krol2022annihilation,krol2022annihilation} by increasing non Hermiticity. However, very few experiments are reported to measure the topological charge associated with it, in parameter space~\cite{gao2015observation}. So our model gives a new way of realizing EP and calculate the \textit{half-integer} topological charge in the parameter space of non-Hermitian couplings. The extended modes have the potential to be highly beneficial in broad-area and efficient laser emission~\cite{longhi2018non} in mirrored resonators as a possible application.\\
\indent{} \textit{The authors are grateful to Henning Schomerus for very insightful discussions.}
 
 \bibliography{paper}

\clearpage

\begin{widetext}

\section{Supplemental Material}

\subsection{CHARACTERISTICS OF SINGLE NON-HERMITIAN SSH CHAIN .} \label{SA}

In this section we discuss the specific similarity transformation \textit{S}~\cite{yao2018edge,li2022dynamic} by which Hamiltonian of single non-Hermitian SSH chain $H_{I/II}$ can be transformed into Hermitian one $\Bar{H}_{I/II}$ through the following form $S^{-1}H_{I/II}S$=$\Bar{H}_{I/II}$. This transformation leaves the eigen values unchanged while the eigen vectors transforms as $\ket{\psi} $ $\rightarrow$ $S^{-1}\ket{\psi}$.

For $\beta$=0, under the OBC the Hamiltonian $H_{I/II}$ in real space takes the following form ,

\begin{equation} \label{sqA1}
H_{I/II}= 
\begin{bmatrix}
  0  & t_{L}^{I/II}& 0 &  0 & 0\\ 
t_{R}^{I/II}  &  0 & t_{2}  & 0 & 0\\ 
0  & t_{2} & 0 & t_{L}^{I/II} & 0\\ 
0  &  0 & t_{R}^{I/II}  & 0 & \dots \\
0  &  0 & 0  & \vdots & \ddots\\
\end{bmatrix} \ .
\tag{SA1}
\end{equation}

Now considering $r$=$\sqrt{\frac{t_{R}^{I/II}}{t_{L}^{I/II}}}$ and choosing $S$=diag$\left( 1,r,r,r^2,r^2,\dotsc,r^{N-1},r^{N-1},r^{N} \right)$, $\Bar{H}_{I/II}$ reads,

\begin{equation} \label{sqA2}
\Bar{H}_{I/II}= 
\begin{bmatrix}
  0 & \sqrt{t_{L}^{I/II}t_{R}^{I/II}}& 0 &  0 & 0\\ 
\sqrt{t_{L}^{I/II}t_{R}^{I/II}}  &  0 & t_{2}  & 0 & 0\\ 
0  & t_{2} & 0 & \sqrt{t_{L}^{I/II}t_{R}^{I/II}} & 0\\ 
0  &  0 & \sqrt{t_{L}^{I/II}t_{R}^{I/II}}  & 0 & \dots \\
0  &  0 & 0  & \vdots & \ddots\\
\end{bmatrix} \ .
\tag{SA2}
\end{equation}

The Bloch Hamiltonian for the chain I and II in presence of gain and loss reads,
\begin{equation} \label{sqA3}
H_{I/II}(k)= 
\begin{bmatrix}
  \pm i\beta &
  t_{L}^{I/II}+t_{2}e^{ik}\\ 
  t_{R}^{I/II}+t_{2}e^{-ik} & \pm i\beta
\end{bmatrix}
\tag{SA3}
\end{equation} 
 with the corresponding eigenenergies,

 \begin{equation} \label{sqA4}
 \epsilon_{1,2}^{I/II}=\pm i\beta\mp\sqrt{t_{2}^2+t_{L}^{I/II}t_{R}^{I/II} + e^{ik}t_{2}t_{R}^{I/II}+ e^{-ik}t_{2}t_{L}^{I/II}},
 \tag{SA4}
\end{equation} 

 right eigenvectors,

\begin{equation} \label{sqA5}
 \ket{\Gamma_{1,2}^{R}}=\bigl\{ \mp\frac{\sqrt{t_{2}^2+t_{L}^{I/II}t_{R}^{I/II} + e^{ik}t_{2}t_{R}^{I/II}+ e^{-ik}t_{2}t_{L}^{I/II}}}{ e^{-ik}t_{2}+t_{R}^{I/II}},1\bigl\}
 \tag{SA5}
\end{equation} 

and associated left eigenvectors,

\begin{equation} \label{sqA6}
 \ket{\Gamma_{1,2}^{L}}=\bigl\{ \mp\frac{\sqrt{t_{2}^2+t_{L}^{I/II}t_{R}^{I/II} + e^{ik}t_{2}t_{L}^{I/II}+ e^{-ik}t_{2}t_{R}^{I/II}}}{ e^{-ik}t_{2}+t_{L}^{I/II}},1\bigl\}.
 \tag{SA6}
\end{equation} 

The associated right and left eigen vectors corresponding to the eigen values $\bigl\{\epsilon_{i},\epsilon_{j}\bigl\}$ satisfies the biorthogonal normalization $\braket{\Gamma{i}^{L}|\Gamma_{j}^{R}}$=$\delta_{ij}$ and the normalization constant given by $N_{i}$=$\frac{1}{\sqrt{\braket{\Gamma_{i}^{L}|\Gamma_{i}^{R}}}}$.

The winding number $W$ for a given band  $\epsilon_{i}$ can be calculated using the following relationship. 
\begin{equation} \label{sqA7}
W_{i}=\frac{1}{\pi}\int_{}^{} \bra{\Gamma_{i}^{L}}\partial_{k} \ket{\Gamma_{i}^{R}} \,dk
\tag{SA7}.
\end{equation} 

However, in the presence of NHSE, the Brillouin zone (BZ) $e^{ik}$ defined in a unit circle by the momentum wave vector $k$ is insufficient to predict the system's bulk topology. This discrepancy is overcome by introducing the concept of generalized Brillouin zone (GBZ) in the followngs Refs.~\cite{yao2018edge,guo2021exact}.
%~\cite{yao2018edge,guo2021exact}.

From (\ref{sqA3}) it is evident that the presence of any constant imaginary onsite potential ($i\beta$) does not change the eigen functions $\ket{\Gamma_{1,2}^{R}}$, $\ket{\Gamma_{1,2}^{L}}$. So the presence of gain/loss does not alternate the characteristics of a single chain when decoupled i.e. $t_{c}$=0 . It is demonstrated in Fig.~\ref{sf1} considering OBC in the presence of $\beta$. We plot all the eigenfunctions of chain I in Fig.~\ref{sf1}(a), which shows the localization of all the states at the left end. The localization of topological modes is shown in the inset of Fig.~\ref{sf1}(a). Because non-reciprocity is reversed in chain II, all eigenfunctions are concentrated in the right open end, as shown in Fig.~\ref{sf1}(b).  This apparent localization of all the eigen modes at one end of the chain is refers the NHSE. In Fig.~.~\ref{sf1}(c) we plot the respective eigenfunctions of the two topological  modes of the Hermitian hamiltonian $\Bar{H}_{I}$. These two modes are positioned at the respective two open ends of the chain satisfying the boundary conditions ($\psi_{1{}B}^{I},\psi_{N_{}A}^{I}$)=0. We plot the band spectrum as a function of the parameter $t_{2}$ in Fig.~\ref{sf1}(d).The gray and red colors illustrate the bulk and edge states, shifted by $\beta$, respectively. Inset of Fig.~\ref{sf1}(d) corresponds to the \textit{W} plot which shows that the value of \textit{W} remains quantized to $1$ as edge modes appear, and that the jump in the value of \textit{W} for certain values of $t_{2}$ corresponds exactly to the merging of the edge modes to the bulk modes, satisfying the generalized (BBC)~\cite{yao2018edge}.

\begin{figure*}[tb]
\setcounter{figure}{0}
\renewcommand{\figurename}{Fig.}
\renewcommand{\thefigure}{S\arabic{figure}}
%\textbf{(a) Eigen energy spectra} 
\begin{center}
\setlength{\tabcolsep}{-.005pt}
\begin{tabular}{c c c c }
(a) & (b) & (c) & (d) 
\\
\includegraphics[width=\z cm]{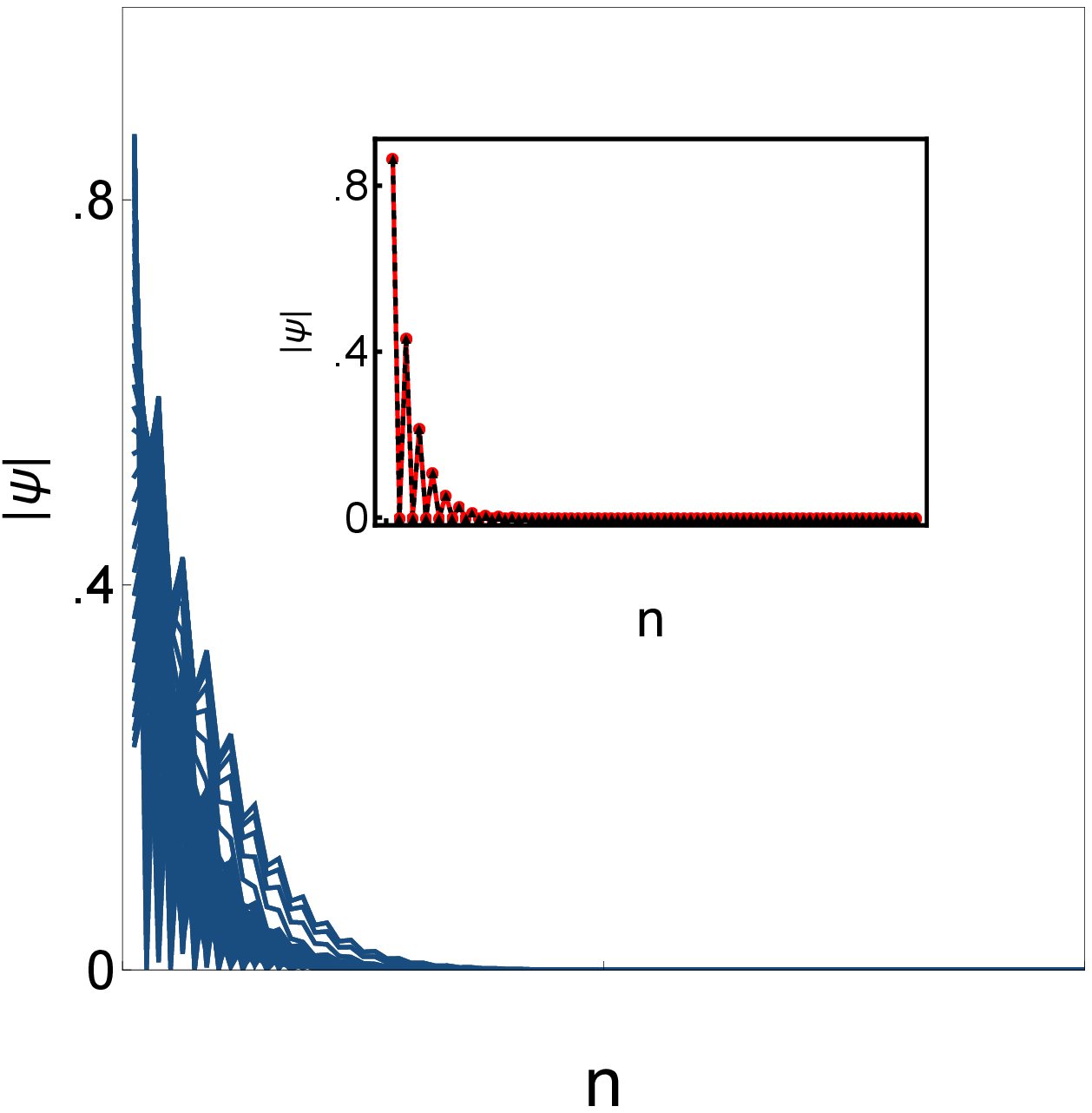}
&  
\includegraphics[width=\z cm]{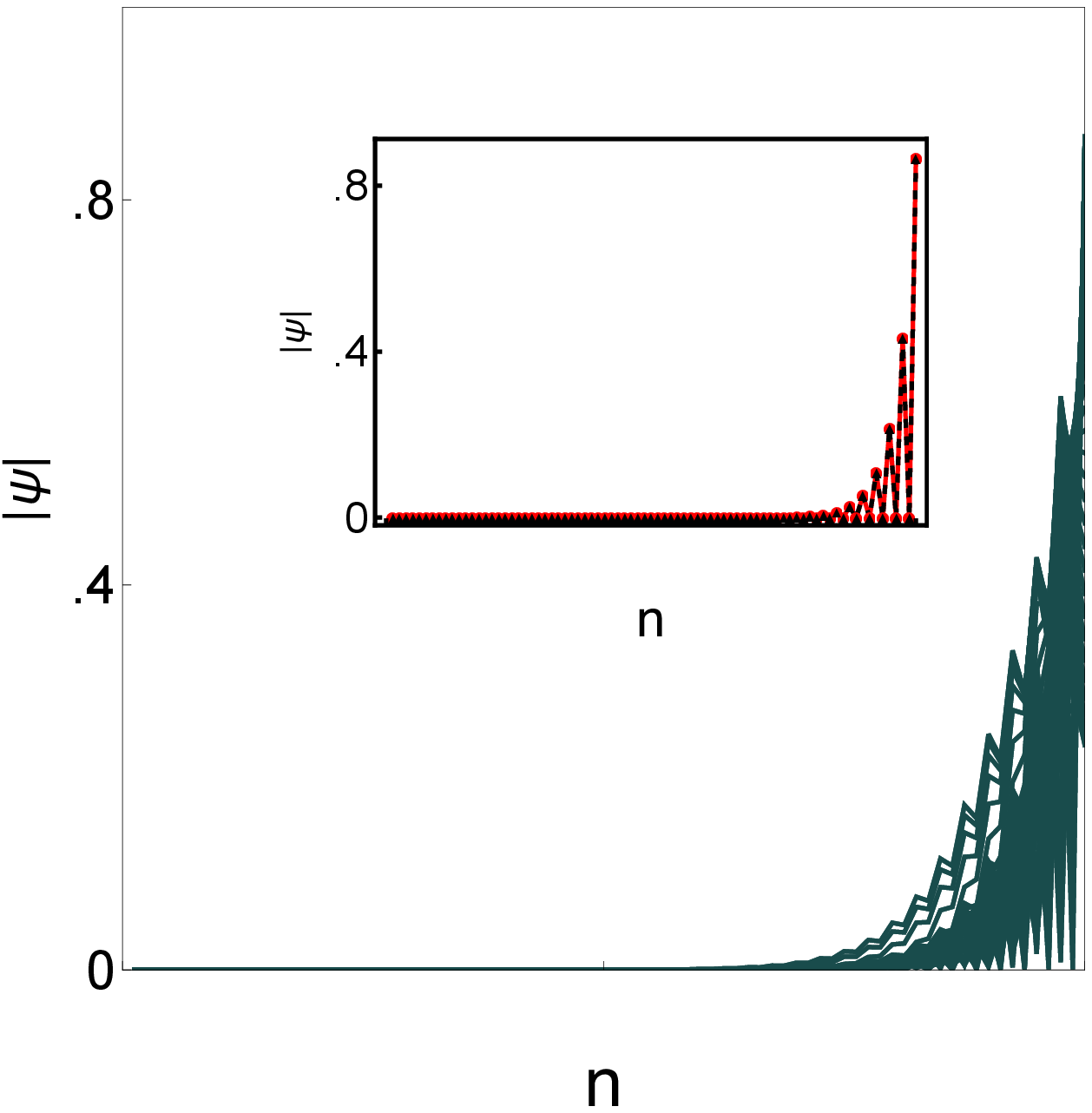} 
&  
\includegraphics[width=\z cm]{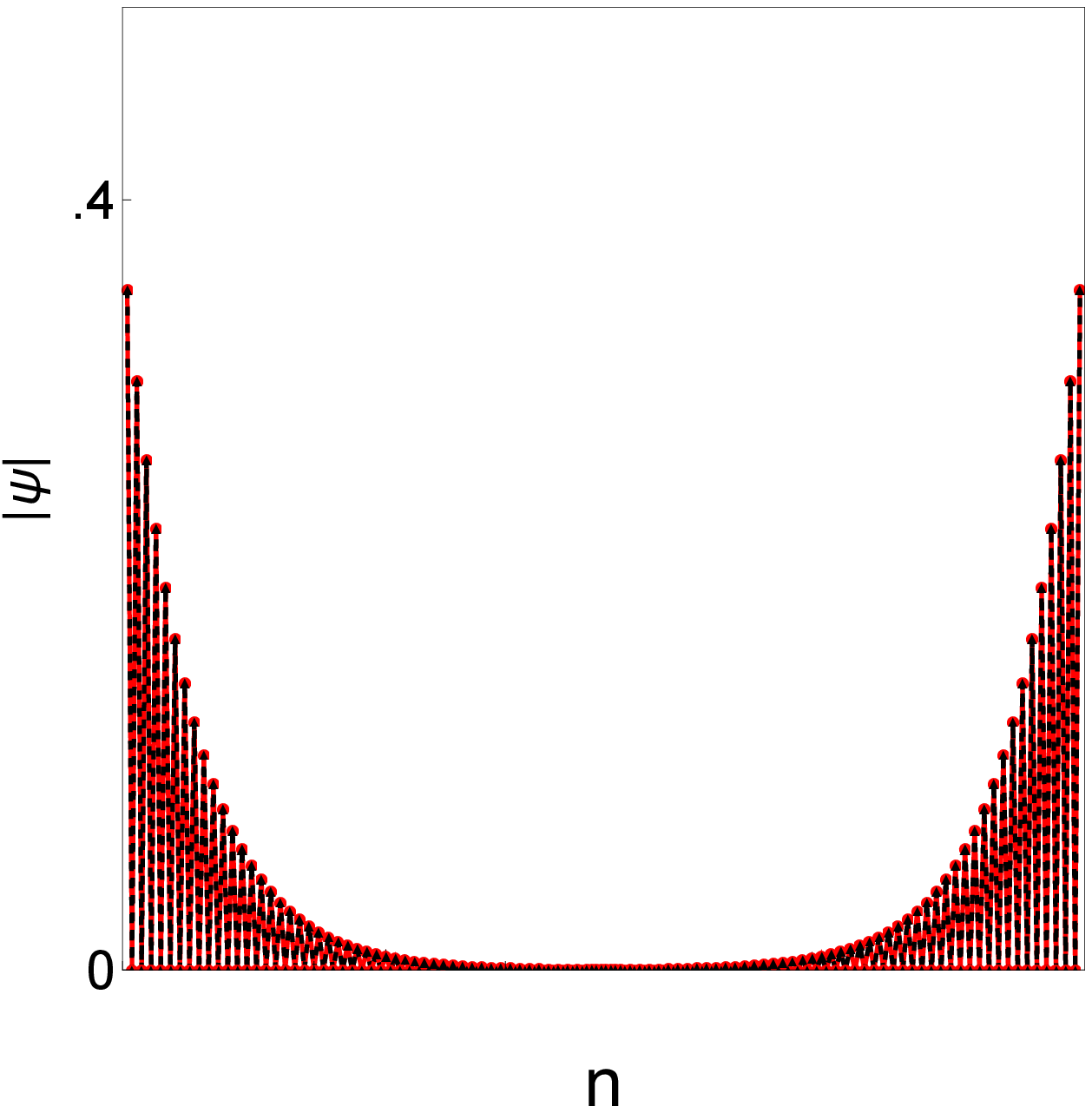} 
&  
\includegraphics[width=\z cm]{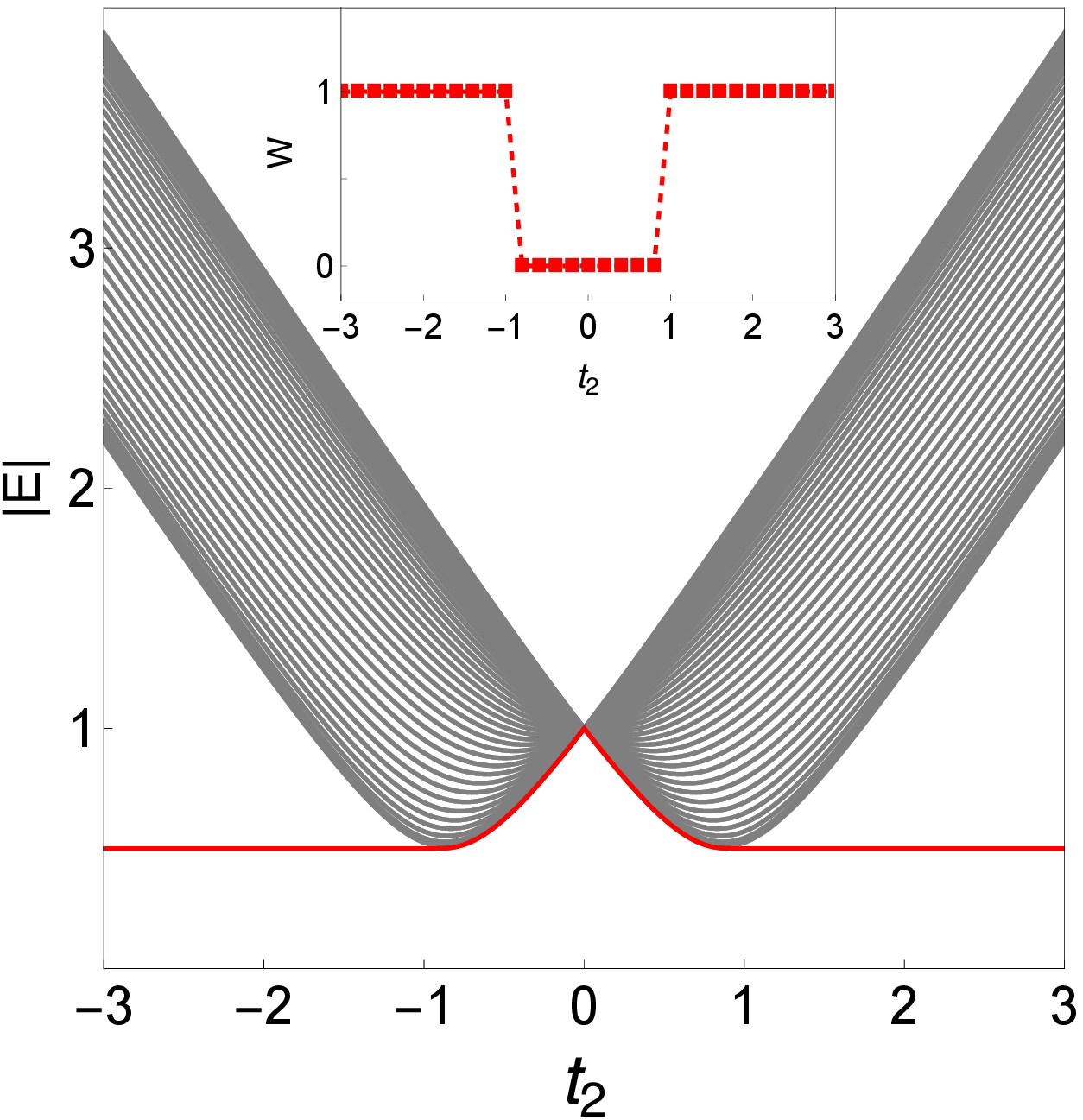} 
\end{tabular}
\end{center} 
\caption{ (a) Eigenfunctions localization in the left end of chain I in OBC (b) Eigenfunctions localization in the right end of chain II in OBC. Inset in (a) and (b) shows the distribution of topological modes. (c) Topological boundary modes obtained from diagonalizing $\Bar{H_{I}}$ in (\ref{sqA2}). (d) Energy spectra $E$ obtained by diagonalizing (\ref{sqA1}) of  single chain with varying $t_{2}$ in OBC. Inset of (d) illustrates Winding number W as function of $t_{2}$.}
\label{sf1}
\end{figure*}

\vspace{0.6 cm}

\subsection{DETAILS FOR ANALYTICAL SOLUTIONS OF EIGEN VALUES AND EIGEN FUNCTIONS OF COUPLED NON HERMITIAN SSH MODEL.} \label{SB}

The details for the analytical solution of the 1D coupled non-Hermitian SSH model, with its Hamiltonian provided by (\ref{eq1}) in the main text, is given here.

 We employ The real-space Schrödinger equation  $H\ket{\Psi}$=$E_{OBC}\ket{\Psi}$  to determine the eigenvalues  $E_{OBC}$ of the open boundary chain. With $\ket{\Psi}$=
 $\left( \psi_{1A}^{I},\psi_{1B}^{I},\dotsc,\psi_{N_{}A}^{I},\psi_{N_{}B}^{I},\psi_{N_{}A}^{II},\psi_{N_{}B}^{II},\dotsc,
 \psi_{1A}^{II},\psi_{1B}^{II} \right)$, we obtain the following recurrance relation for the eigen function within the bulk of chain I,
 \begin{align} \label{sq1}
 \begin{cases}
    %\begin{align}
     E_{OBC}^{I} \psi_{n+1A}^{I}=t_{L}^{I}\psi_{n+1B}^{I}+t_{2}\psi_{n B}^{I}  \\
     E_{OBC}^{I}\psi_{nB}^{I}=t_{R}^{I}\psi_{nA}^{I}+t_{2}\psi_{n+1A}^{I}.
    %\end{align}
    \end{cases}
    \tag{SB1}
\end{align}
Similarly for the chain II we obtain,
\begin{align} \label{sq2}
 \begin{cases}
    %\begin{align}
     E_{OBC}^{II} \psi_{nA}^{II}=t_{L}^{II}\psi_{nB}^{II}+t_{2}\psi_{n+1B}^{II}  \\
     E_{OBC}^{II}\psi_{n+1B}^{II}=t_{R}^{II}\psi_{n+1A}^{I}+t_{2}\psi_{nA}^{II}.
    %\end{align}
    \end{cases}
    \tag{SB2}
\end{align}
where $E_{OBC}^{I}$= $E_{OBC}^{}$-$i\beta$, $E_{OBC}^{II}$= $E_{OBC}^{}$+$i\beta$ and $n$ is the unit cell index . We can consider an ansatz for the eigenfunctions of both chains as a linear combination~\cite{yao2018edge,guo2021exact} in accordance with the theory of linear difference equations:

\begin{equation}\label{sq3}
    \begin{bmatrix} 
      \psi_{nA}^{I/II} \\ 
      \psi_{nB}^{I/II} 
     \end{bmatrix}= (\lambda_{1}^{I/II})^{n} \begin{bmatrix} 
                \phi_{A1}^{I/II} \\ 
                \phi_{B1}^{I/II} 
               \end{bmatrix} + (\lambda_{2}^{I/II})^{n} \begin{bmatrix}
                \phi_{A2}^{I/II} \\ 
                \phi_{B2}^{I/II} 
               \end{bmatrix} ; 
               \tag{SB3}
\end{equation}

 Substituting (\ref{sq3}) in (\ref{sq1}) and (\ref{sq2}) we got
  \begin{equation}\label{sq4}
 \begin{array}{l}
    \Phi_{Aj}^{I}=\frac{E_{OBC}^{I}\Phi_{Bj}^{I}}{t_{R}^{I}+t_{2}\lambda_{j}^{I}}=\frac{\Big[t_{L}^{I}+\frac{t_{2}}{\lambda_{j}^{I}}\Big]\Phi_{Bj}^{I}}{E_{OBC}^{I}},
     
     \\

      \Phi_{Bj}^{II}=\frac{E_{OBC}^{II}\Phi_{Aj}^{II}}{t_{L}^{II}+t_{2}\lambda_{j}^{II}}=\frac{\Big[t_{R}^{II}+\frac{t_{2}}{\lambda_{j}^{II}}\Big]\Phi_{Aj}^{II}}{E_{OBC}^{II}},

     \end{array}
     \tag{SB4}
 \end{equation}

 with j=1,2. From (\ref{sq4})  we got the following expressions for $\lambda_{1/2}^{I/II}$,

 \begin{equation}\label{sq5}
 \begin{array}{l}
     
     \lambda_{1/2}^{I}=\frac{{E_{OBC}^{I}}^2-t_{R}^{I}t_{L}^{I}-t_{2}^2\pm\sqrt{\left( {E_{OBC}^{I}}^2-t_{R}^{I}t_{L}^{I}-t_{2}^2 \right){}^2-4 t_2^2t_{R}^{I}t_{L}^{I}}}{2 t_2 t_{L}^{I}},
      \\
    \lambda_{1/2}^{II}=\frac{{E_{OBC}^{II}}^2-t_{R}^{II}t_{L}^{II}-t_{2}^2\pm\sqrt{\left( {E_{OBC}^{II}}^2-t_{R}^{II}t_{L}^{II}-t_{2}^2 \right){}^2-4 t_2^2t_{R}^{II}t_{L}^{II}}}{2 t_2. t_{R}^{II}} 
\end{array}
\tag{SB5}
 \end{equation}

$\lambda_{1,2}^{I,II}$ denotes the generalized Brillouin zone (GBZ)~\cite{yao2018edge,guo2021exact} of the bulk bands. In Fig.~\ref{sf2}, we plot the GBZ points for all bands for different values of $\beta$.

As illustrated in Fig.~\ref{f1}(a) in the main text, the boundary equations for the eigen functions in the real space  can be expressed as follows:

 \begin{align} \label{sq6}
 \begin{cases}
    %\begin{align}
     E_{OBC}^{I} \psi_{1A}^{I}=t_{L}^{I}\psi_{1B}^{I},  \\
     E_{OBC}^{I}\psi_{NB}^{I}=t_{R}^{I}\psi_{NB}^{I}+t_{2}\psi_{NA}^{II},      \\
     E_{OBC}^{II}\psi_{NA}^{II}=t_{L}^{II}\psi_{NB}^{II}+t_{2}\psi_{NB}^{I},      \\
     E_{OBC}^{II}\psi_{1B}^{II}=t_{R}^{II}\psi_{1A}^{II}.
    %\end{align}
    \end{cases}
    \tag{SB6}
\end{align}

The ansatz in (\ref{sq3}) should satisfy the boundary conditions (\ref{sq6}). Now with the substitution of (\ref{sq4}),  (\ref{sq6}) is written in terms of the coefficients $\bigl\{  \Phi_{(A/B)j}^{I/II} \bigl\}$ with (j=1,2). Additionally, we can obtain the coupled equations in (\ref{sq6}) including only the set $\Theta$ as $H_{B}\Theta$=0, where  $\Theta$= $\left[\phi_{B1}^{I},\phi_{B2}^{I},\phi_{A1}^{II},\phi_{A2}^{II}\right]$. $H_{B}$ has the following matrix form:,

  \begin{equation}\label{sq7}
 H_{B}=\begin{bmatrix}
t_{2} & t_{2} & 0 & 0 \\
t_{2} {\lambda_{1}^{I}}^{N+1}\eta_{1}^{I} & t_{2} {\lambda_{2}^{I}}^{N+1}\eta_{2}^{I} & -t_{2} {\lambda_{1}^{II}}^{N} & -t_{2} {\lambda_{2}^{II}}^{N} \\
 -t_{2} {\lambda_{1}^{I}}^{N} &  -t_{2} {\lambda_{2}^{I}}^{N} & t_{2} {\lambda_{1}^{II}}^{N+1}\eta_{1}^{II} & t_{2} {\lambda_{2}^{II}}^{N+1}\eta_{2}^{II} \\
0 & 0 & t_{2} & t_{2}
\end{bmatrix},
\tag{SB7}
\end{equation}
 where $\eta_{j}^{I}$=$\frac{E_{OBC}^{I}}{t_{R}^{I}+t_{2}\lambda_{j}^{I}}$ and $\eta_{j}^{II}$=$\frac{E_{OBC}^{II}}{t_{L}^{II}+t_{2}\lambda_{j}^{II}}$.

 \begin{figure*}[tb]
\setcounter{figure}{0}
\renewcommand{\figurename}{Fig.}
\renewcommand{\thefigure}{S2}
%\textbf{(a) Eigen energy spectra} 
\begin{center}
\setlength{\tabcolsep}{-.0005pt}
\begin{tabular}{c c c c}

\textbf{$\beta$=0.2$\gamma$} & \textbf{$\beta$=0.6$\gamma$} & \textbf{$\beta$=$\gamma$} & \textbf{$\beta$=1.4$\gamma$}
\\
\includegraphics[width=4.0 cm]{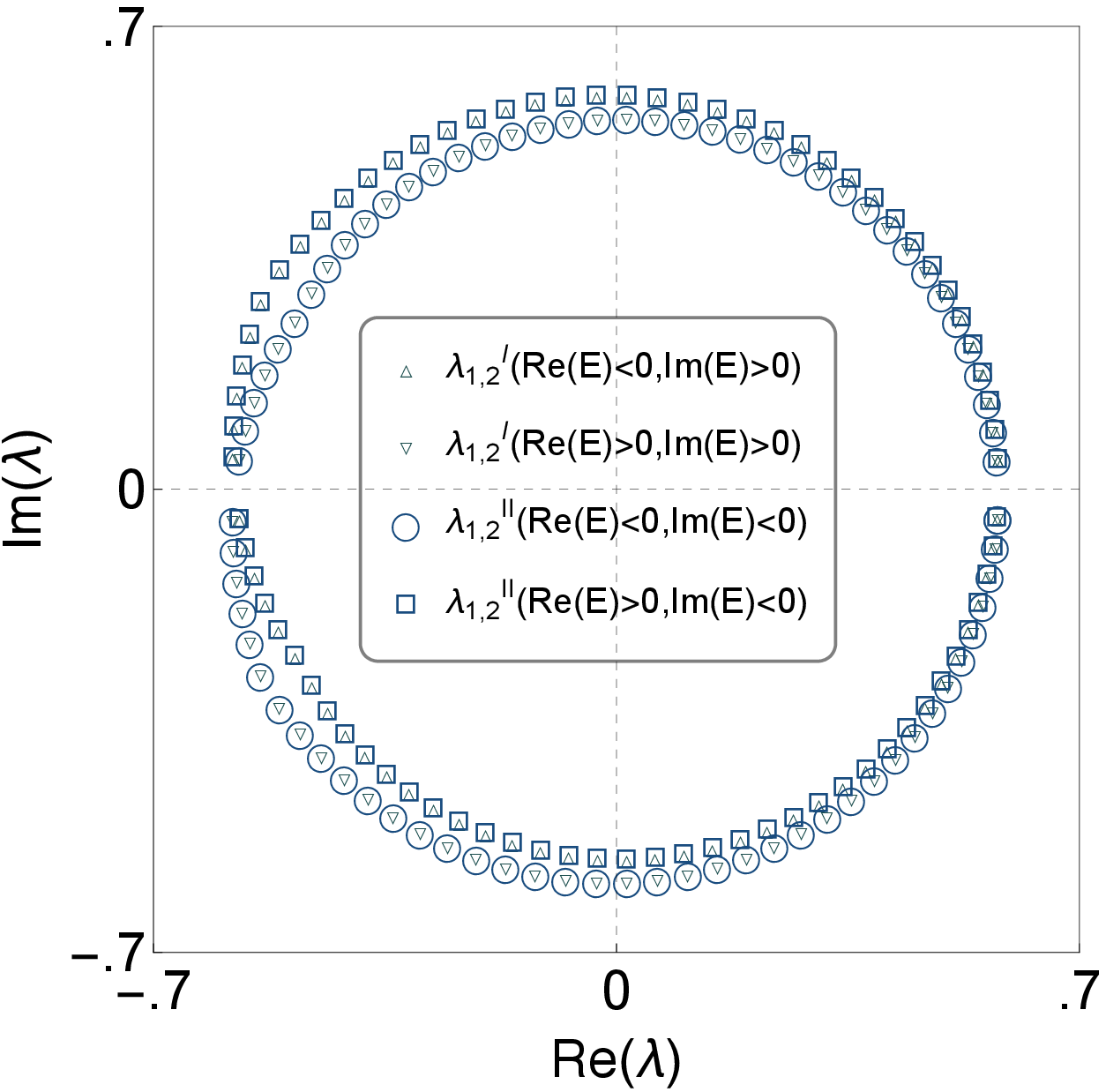}
&  
\includegraphics[width=4.0 cm]{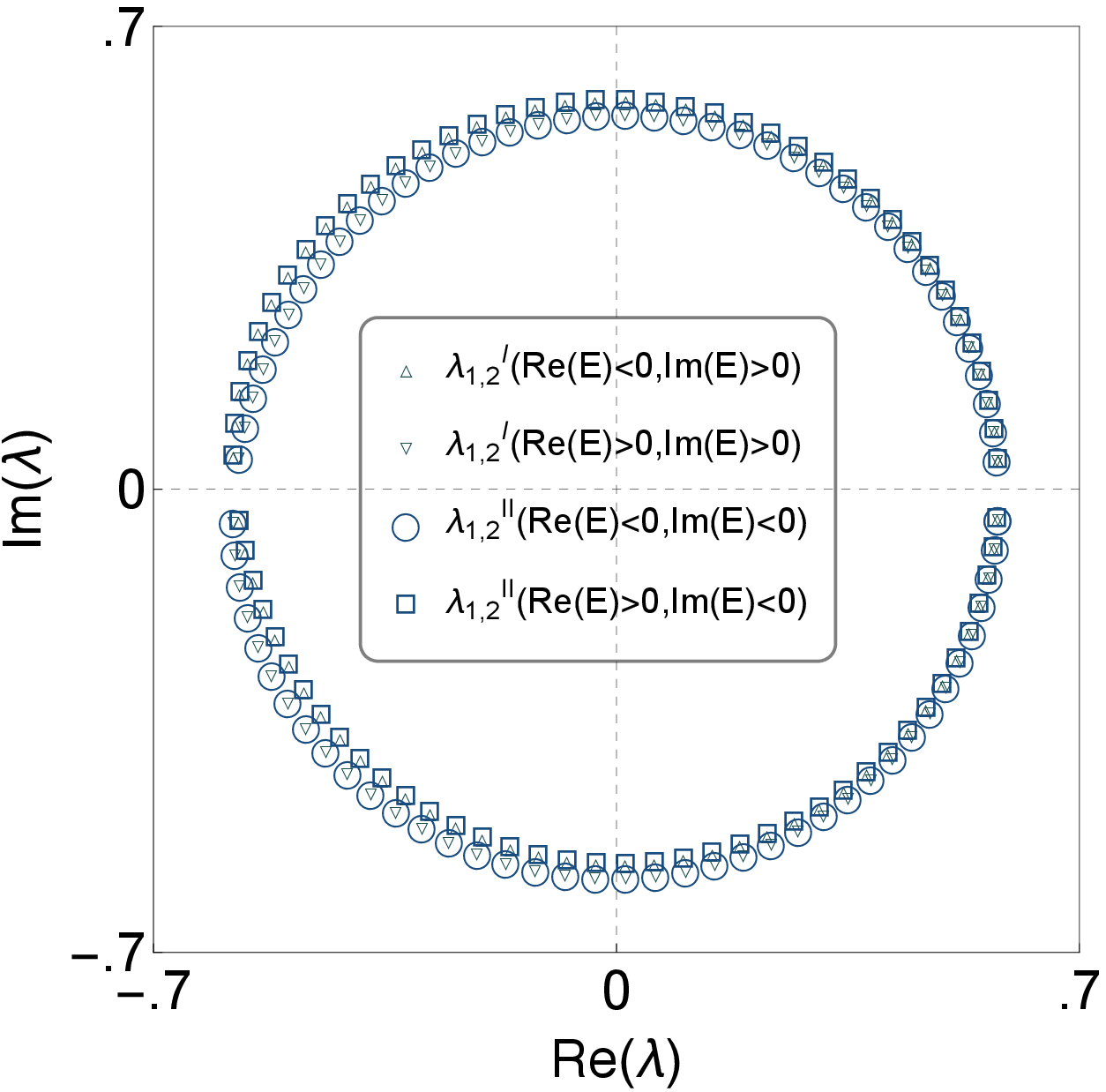} 
&  
\includegraphics[width=4.0 cm]{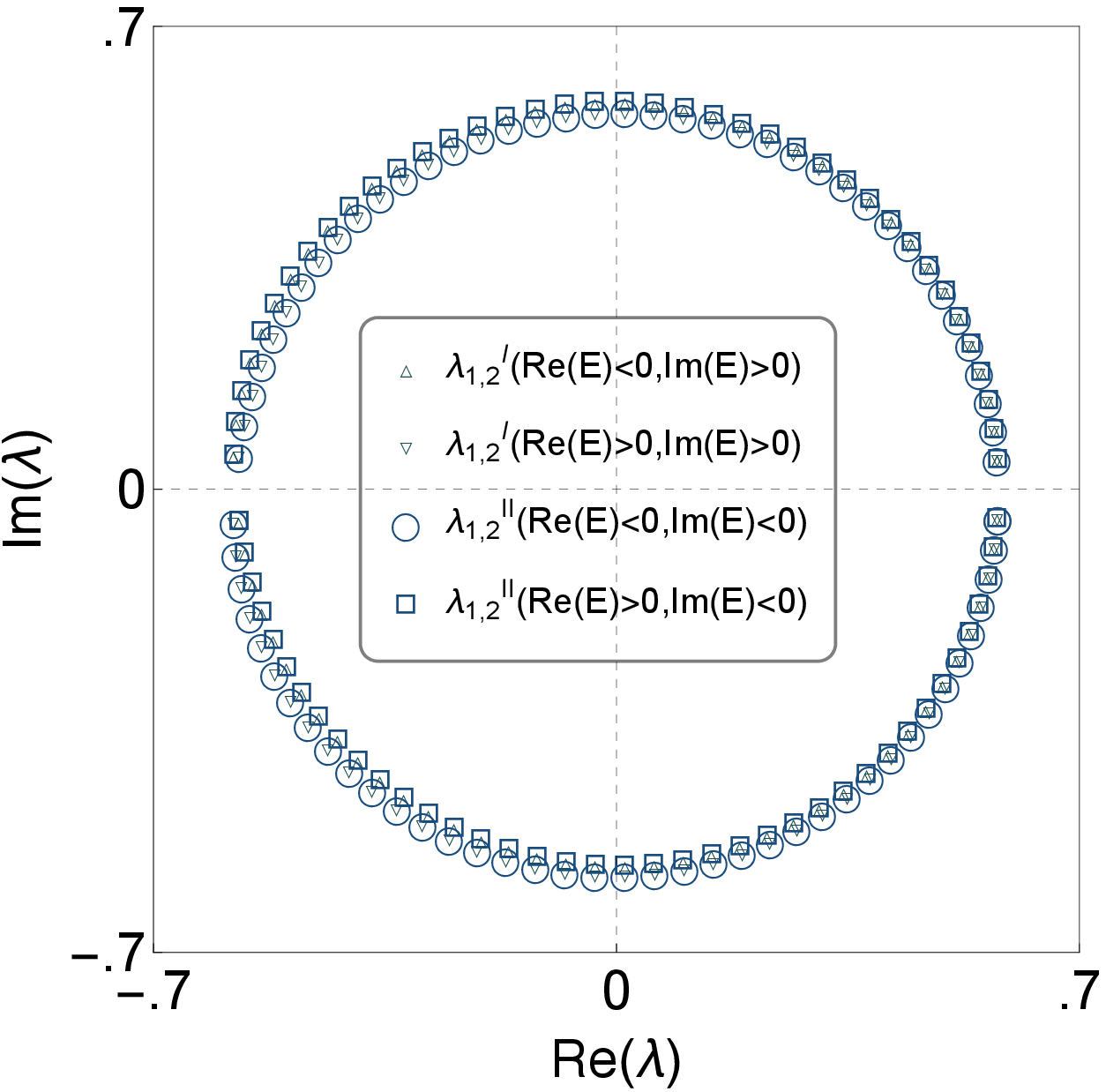} 
&  
\includegraphics[width=4.0 cm]{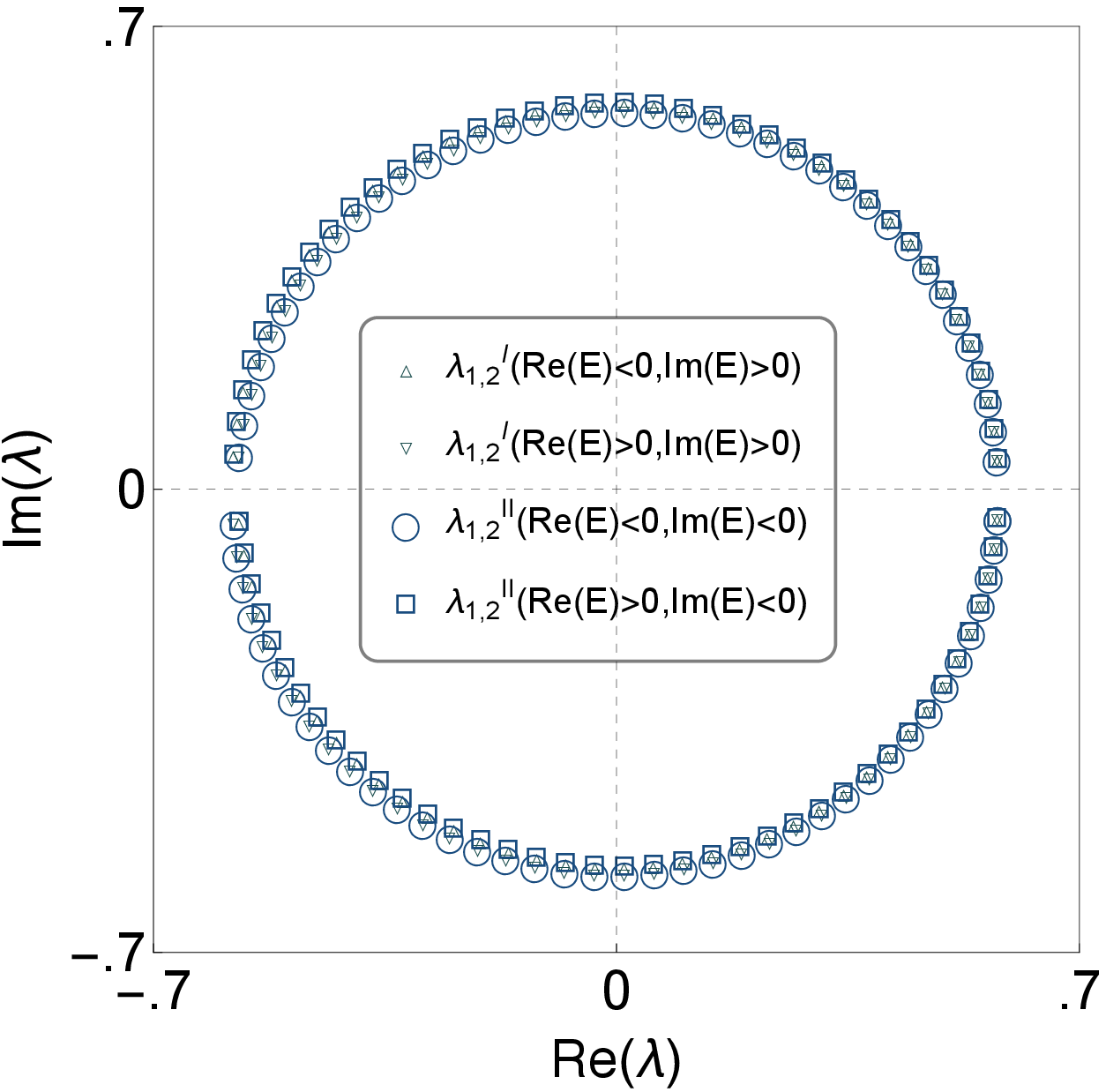} 
\end{tabular}
\end{center} 
\caption{ $\lambda_{1,2}^{I}$,$\lambda_{1,2}^{II}$ for all bulk energy bands as shown in Fig.\ref{f1}(d) of the main text.}
\label{sf2}
\end{figure*}

There exists a nontrivial solution corresponding to coefficients $\Theta$ having nonzero values, as given by $det{H_{B}}$=0. This results to the following characteristic equation,

\begin{equation} \label{sq8}
\begin{split}
&{\lambda_{1}^{I}}^{N}\left({ {\lambda_{2}^{II}}^{N} -\lambda_{1}^{II}}^{N}\right)+{\lambda_{2}^{I}}^{N}\left({\lambda_{1}^{II}}^{N} - {\lambda_{2}^{II}}^{N} \right)
    \\&+\eta_{1}^{I}{\lambda_{1}^{I}}^{N+1}\left({\eta_{1}^{II}\lambda_{1}^{II}}^{N+1} - \eta_{2}^{II}{\lambda_{2}^{II}}^{N+1} \right)
    \\&+\eta_{2}^{I}{\lambda_{2}^{I}}^{N+1}\left({\eta_{2}^{II}\lambda_{2}^{II}}^{N+1} - \eta_{2}^{II}{\lambda_{1}^{II}}^{N+1} \right)=0. 
    \end{split}
    \tag{SB8}
\end{equation}

 When $\lambda_{1,2}^{I,II}(E_{OBC})$ in (\ref{sq5}) is substituted, (\ref{sq8}) becomes the function of N and $E_{OBC}$. The energy eigenvalues in of the finite coupled chain in OBC are thus provided by the set of solutions of the polynomial equation (\ref{sq8}) for a fixed chain length N on both sides. Now assuming that the system size is very large, i.e., N $\gg$ 1 and taking into account that eigenvalues with only real or imaginary value satisfy $|\lambda_{1}^{I/II}|<|\lambda_{2}^{I/II}|$, and we can approximate (\ref{sq8}) by ignoring all the other terms and including only  $({\lambda_{2}^{I}\lambda_{2}^{II}})^{N} (\eta_{1}^{I}\eta_{2}^{II}\lambda_{2}^{I}\lambda_{2}^{II}-1)$. The analytical equation for the \textit{in-gap} states with eigenvalues $\pm \Delta$ is obtained by substituting the coefficient  $(\eta_{1}^{I}\eta_{2}^{II}\lambda_{2}^{I}\lambda_{2}^{II}-1)=0$, which results in
 
\begin{equation} \label{sq9}
\Delta= \pm i \sqrt{t^2-t_{2}^2+\beta^2-\gamma^2}.
\tag{SB9}
\end{equation}
%The other two \textit{in-gap} states of energies +$i\beta$ and -$i\beta$  can be obtained from the constraint  $\eta_{1}^{I}\eta_{2}^{II}=0$ or  $\eta_{2}^{I}\eta_{1}^{II}=0$.

Using (\ref{sq1}) and (\ref{sq6}), we can now derive the following relations for the chain I.

\begin{align} \label{sq10}
 \begin{cases}
    %\begin{align}
     \Phi_{B1}^{I}=\frac{\Phi_{A1}^{I}}{\eta_{1}^{I}},  \\
     \Phi_{B2}^{I}= \Phi_{B1}^{I}\frac{ \lambda_{1}^{I}}{ \lambda_{2}^{I}}\frac{E_{OBC}^{I}\eta_{1}^{I}-t_{L}^{I}}{-E_{OBC}^{I}\eta_{2}^{I}+t_{L}^{I}} ,      \\
      \Phi_{A2}^{I}=\Phi_{B2}^{I}\eta_{2}^{I}.
    %\end{align}
    \end{cases}
     \tag{SB10}
\end{align}

Similarly, using (\ref{sq2}) and (\ref{sq6}), we can now derive the following relations for the chain II.

\begin{align} \label{sq11}  \tag{SB11}
 \begin{cases}
    %\begin{align}
     \Phi_{A1}^{II}=\frac{\Phi_{B1}^{II}}{\eta_{1}^{II}},  \\
     \Phi_{A2}^{II}= \Phi_{A1}^{II}\frac{ \lambda_{1}^{II}}{ \lambda_{2}^{II}}\frac{E_{OBC}^{II}\eta_{1}^{II}-t_{R}^{II}}{-E_{OBC}^{II}\eta_{2}^{II}+t_{R}^{II}} ,      \\
      \Phi_{B2}^{I}=\Phi_{A2}^{II}\eta_{2}^{II}.
    %\end{align}
    \end{cases}
    \end{align}

The following expression for the eigen function with eigen energy $E_{OBC}$ in real space is obtained by substituting (\ref{sq10}) and (\ref{sq11}) in (\ref{sq3}).

 \begin{figure*}[tb]
\setcounter{figure}{0}
\renewcommand{\figurename}{Fig.}
\renewcommand{\thefigure}{S3}
%\textbf{(a) Eigen energy spectra} 
\begin{center}
\setlength{\tabcolsep}{-.0005pt}
\begin{tabular}{c c c c}

\textbf{$\beta$=0.2$\gamma$} & \textbf{$\beta$=0.6$\gamma$} & \textbf{$\beta$=$\gamma$} & \textbf{$\beta$=1.4$\gamma$} 
\\
\includegraphics[width=4.0 cm]{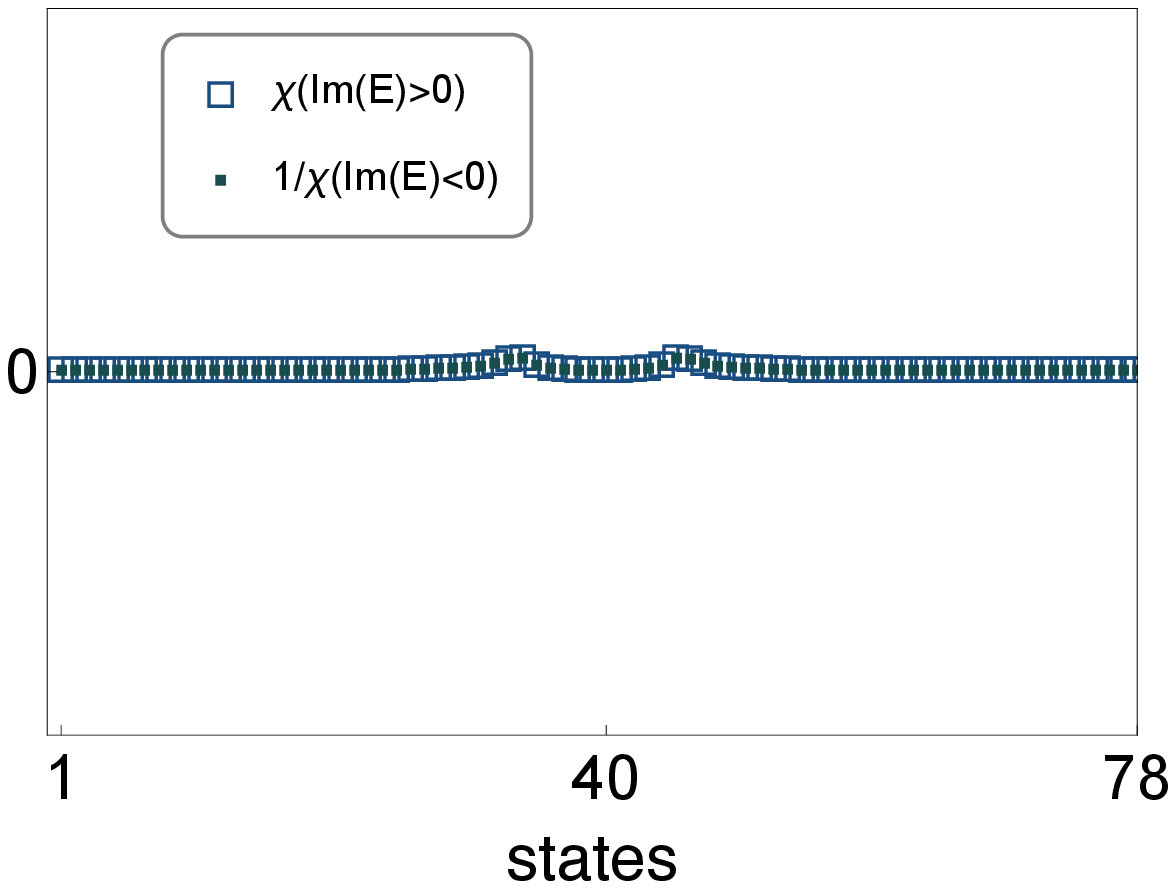}
&  
\includegraphics[width=4.0 cm]{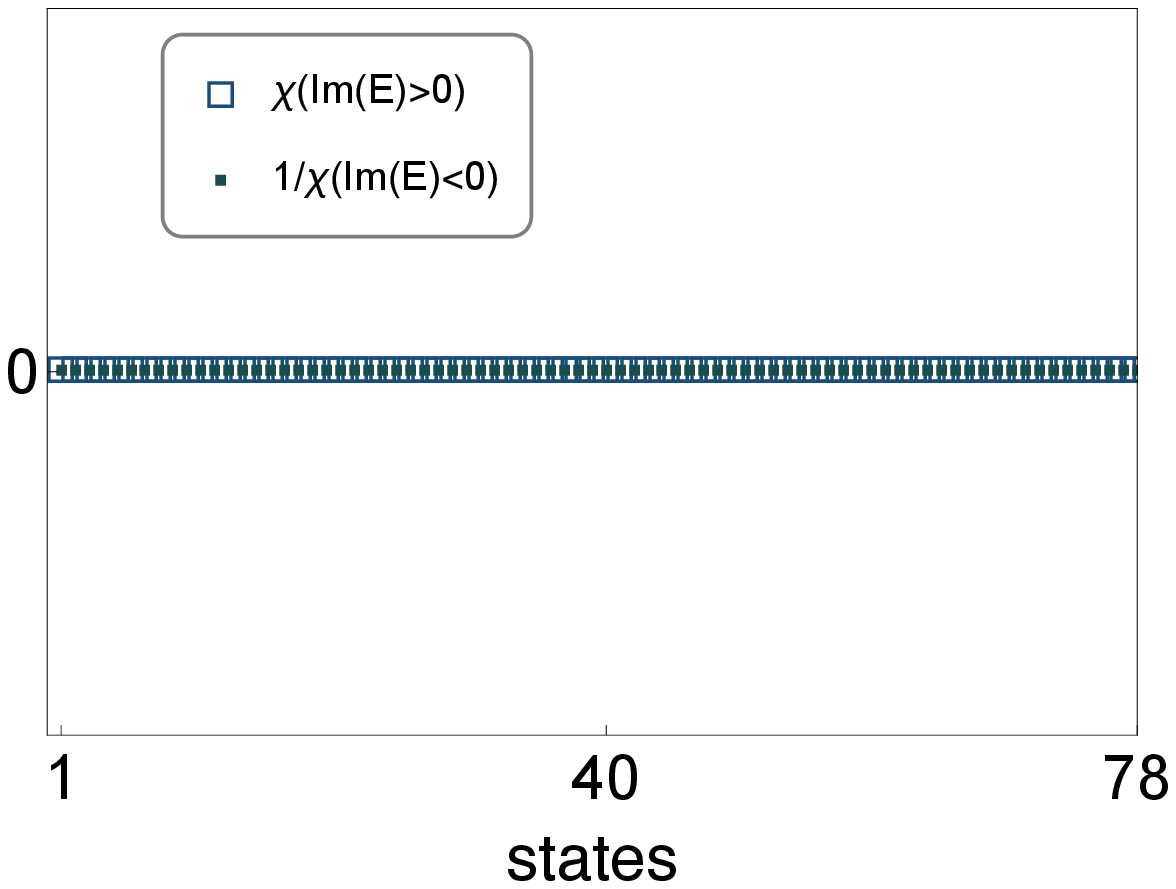} 
&  
\includegraphics[width=4.0 cm]{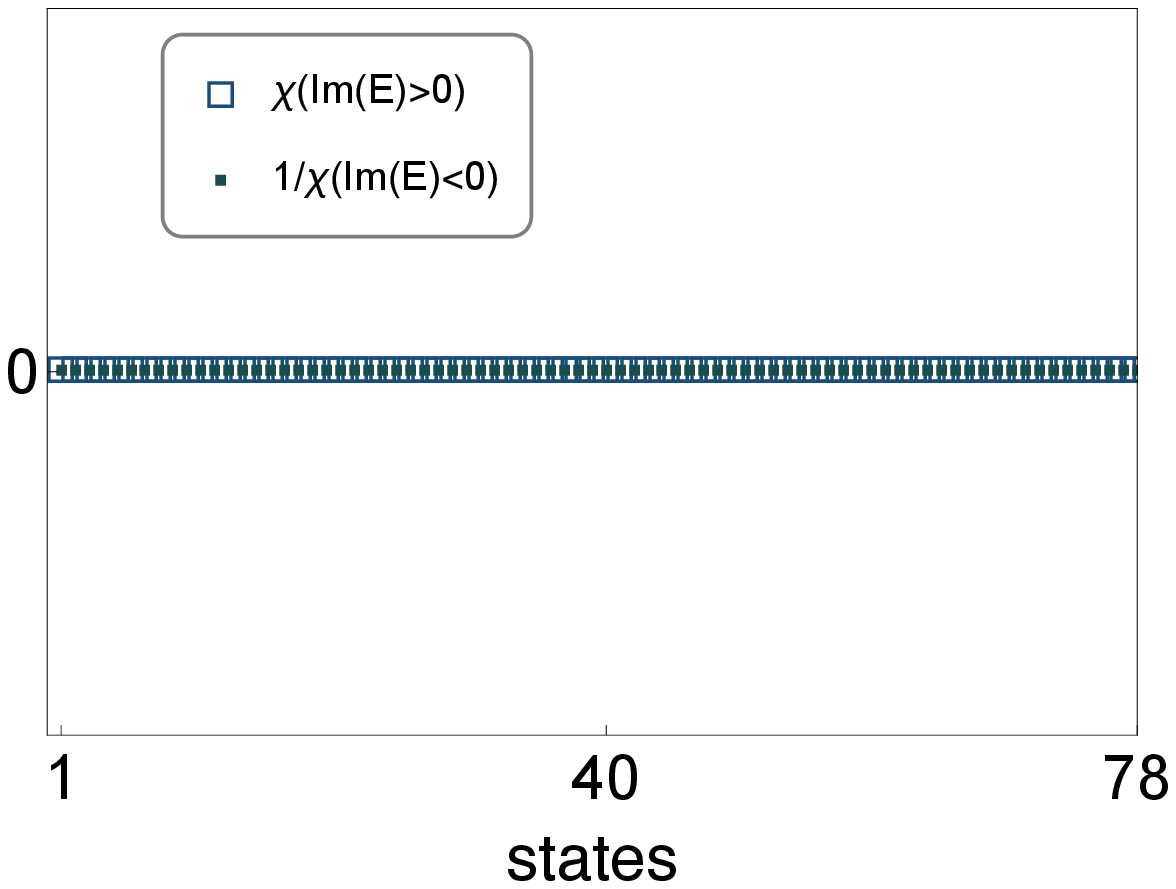} 
&  
\includegraphics[width=4.0 cm]{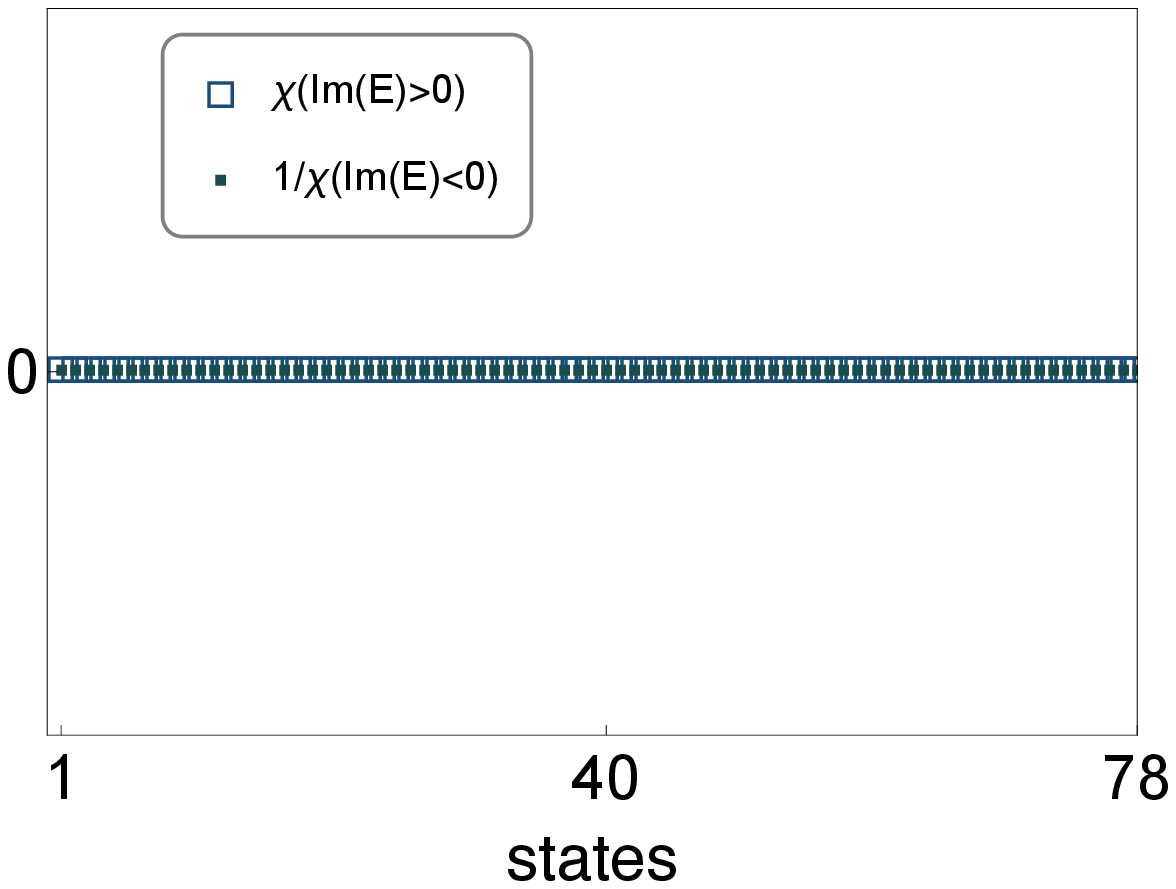} 
\end{tabular}
\end{center} 
\caption{ Illustration of $\chi$($E_{OBC}$) from (\ref{sq13}) for the bulk energy bands as shown in Fig.\ref{f1}(d) of the main text.}
\label{sf3}
\end{figure*}

%\begin{widetext}
\begin{align}\label{sq12} \tag{SB12}
%\begin{array}{l}
\psi_{nA}^{I}&= \Bigg[\frac{\left({E_{OBC}^{I}}^2 +t_{L}^{I}t_{R}^{I} -t_{2}^2 +\delta_{I} \right) \left( \frac{{E_{OBC}^{I}}^2 -t_{L}^{I}t_{R}^{I} -t_{2}^2 -\delta_{I}}{t_2 t_{L}^{I}}\right)^n}{{-E_{OBC}^{I}}^2 -t_{L}^{I}t_{R}^{I} +t_{2}^2 +\delta_{I}}+\left(\frac{{E_{OBC}^{I}}^2 -t_{L}^{I}t_{R}^{I} -t_{2}^2 +\delta_{I}}{t_2 t_{L}^{I}}\right)^n\Bigg]\frac{\Phi_{A1}^{I}}{2^{n}},  \nonumber\\
 \psi_{nB}^{I}&= \frac{\Phi_{A1}^{I}E_{OBC}^{I}t_{R}^{I}}{2^{n-1}} \Bigg[\frac{\left( \frac{{E_{OBC}^{I}}^2 -t_{L}^{I}t_{R}^{I} -t_{2}^2 +\delta_{I}}{t_2 t_{L}^{I}}\right)^n + \left( \frac{{E_{OBC}^{I}}^2 -t_{L}^{I}t_{R}^{I} -t_{2}^2 -\delta_{I}}{t_2 t_{L}^{I}}\right)^n}{{E_{OBC}^{I}}^2 +t_{L}^{I}t_{R}^{I} -t_{2}^2 +\delta_{I}}\Bigg] ,\nonumber \\
 \psi_{nB}^{II}&= \Bigg[\frac{\left({E_{OBC}^{II}}^2 -t_{L}^{II}t_{R}^{II} -t_{2}^2 +\delta_{II} \right) \left( \frac{{E_{OBC}^{II}}^2 -t_{L}^{II}t_{R}^{II} -t_{2}^2 -\delta_{II}}{t_2 t_{R}^{II}}\right)^n}{{-E_{OBC}^{II}}^2 -t_{L}^{II}t_{R}^{II} +t_{2}^2 +\delta_{II}} + \left(\frac{{E_{OBC}^{II}}^2 -t_{L}^{II}t_{R}^{II} -t_{2}^2 +\delta_{II}}{t_2 t_{R}^{II}}\right)^n\Bigg]\frac{\Phi_{B1}^{II}}{2^{n}},  \nonumber\\
 \psi_{nA}^{II}&= \frac{\Phi_{B1}^{II}E_{OBC}^{II}t_{L}^{II}}{2^{n-1}} \Bigg[\frac{\left( \frac{{E_{OBC}^{II}}^2 -t_{L}^{II}t_{R}^{II} -t_{2}^2 +\delta_{II}}{t_2 t_{R}^{II}}\right)^n + \left( \frac{{E_{OBC}^{II}}^2 -t_{L}^{II}t_{R}^{II} -t_{2}^2 -\delta_{II}}{t_2 t_{R}^{II}}\right)^n}{{E_{OBC}^{II}}^2 +t_{L}^{II}t_{R}^{II} -t_{2}^2 +\delta_{II}}\Bigg], \nonumber
 %\nonumber\\
 %&-\frac{\left( \frac{{E_{OBC}^{I}}^2 -t_{L}^{I}t_{R}^{I} -t_{2}^2 -\sqrt{\left(-{E_{OBC}^{I}}^2+t_{L}^{I}t_{R}^{I}+t_{2}^2\right)^2-4t_{L}^{I}t_{R}^{I} t_{2}^2}}{t_2 t_{L}^{I}}\right)^n}{{E_{OBC}^{I}}^2 +t_{L}^{I}t_{R}^{I} -t_{2}^2 +\sqrt{\left(-{E_{OBC}^{I}}^2+t_{L}^{I}t_{R}^{I}-t_{2}^2\right)^2-4t_{L}^{I}t_{R}^{I} t_{2}^2}} \Bigg]
 %\end{array}
\end{align}
 
%\end{widetext}

where $\delta_{I}=\sqrt{\left(t_{L}^{I}t_{R}^{I}+t_{2}^2-{E_{OBC}^{I}}^2\right)^2-4t_{L}^{I}t_{R}^{I} t_{2}^2}$ and 
$\delta_{II}=\sqrt{\left(t_{L}^{II}t_{R}^{II}+t_{2}^2-{E_{OBC}^{II}}^2\right)^2-4t_{L}^{II}t_{R}^{II} t_{2}^2}$.

 Note that from (\ref{sq12})  we found, $\Psi_{n(A,B)}^{I}$ depends on $\phi_{A1}^{I}$ and $\Psi_{n(A,B)}^{II}$ depends on $\phi_{B1}^{II}$. Using (\ref{sq6}) we can write down a connecting equation between $\phi_{A1}^{I}$ and $\phi_{B1}^{II}$ as,

%\begin{widetext}
\begin{align}\label{sq13}  \tag{SB13}
\Phi_{B1}^{II}=\Phi_{A1}^{I}\frac{\eta_{1}^{II} \lambda_{2}^{II} (E_{OBC}^{II} \eta_{2}^{II}-t_{R}^{II}) \left(\lambda_{1}^{I} {\lambda_{1}^{I}}^{N} (E_{OBC}^{I} \eta_{1}^{I}-t_{L}^{I}) (E_{OBC}^{I}-\eta_{2}^{I} t_{R}^{I})-\lambda_{2}^{I} {\lambda_{1}^{I}}^{N} (E_{OBC}^{I}-\eta_{1}^{I} t_{R}^{I}) (E_{OBC}^{I} \eta_{2}^{I}-t_{L}^{I})\right)}{\lambda_{2}^{I} \eta_{1}^{I} t_{2} (E_{OBC}^{I} \eta_{2}^{I}-t_{L}^{I}) \left(\lambda_{1}^{II} {\lambda_{2}^{II}}^{N} (E_{OBC}^{II} \eta_{1}^{II}-t_{R}^{II})-\lambda_{2}^{II} {\lambda_{1}^{II}}^{N} (E_{OBC}^{II} \eta_{2}^{II}-t_{R}^{II})\right)},
\end{align}
%\end{widetext}

In Fig.~\ref{sf3} we plot $\chi$=$|\phi_{B1}^{II }/\phi_{A1}^{I}|$, which determines the projection of any eigenfunction onto the chain II.
We found that for bulk states with $\operatorname{Im}(E)$$>0$, $|\chi|$ $\to$ 0 demonstrating that these specific bulk states remain in chain I. These specific states also satisfy  $|\lambda_{1}^{I}|\approx|\lambda_{2}^{I}|<1$. Hence, in the limit $n\gg 1$, $|\psi_{n(A/B)}^{I}|$ $\to$ 0 implies localization at the open left end, as seen in Fig.~\ref{f1}(d) of the Main text. Similarly, the localization of the bulk modes $\operatorname{Im}(E)$$<$0 at the open right end can be stems from the conditions $|1/\chi|$ $\to$ 0 and $|\lambda_{1}^{II}|\approx|\lambda_{2}^{II}|<1$. Our analytical formulation therefore agrees with the numerical results.

 The topological discrete mode with energy $E_{OBC}$=i$\beta$ satisfies $E_{OBC}^{I}$=0, $E_{OBC}^{II}$=2i$\beta$ and $|\chi(i\beta)|$ $\to$ 0. Which conclude that this specific mode is solely belongs to the chain $I$, and its localization at the left boundary can be verified from its eigen function in real space
\begin{align} \label{sq14} \tag{SB14}
 \begin{cases}
   \psi_{nA}^{I}(i\beta)= \Bigg[\left( \frac{-t_{R}^{I}}{t_{2}} \right)^n\Bigg]\Phi_{A1}^{I},  \\
    \psi_{nB}^{I}(i\beta)=  0.
    \end{cases}
\end{align}
Similarly the other mode with energy $E_{OBC}$=-i$\beta$ satisfies $E_{OBC}^{I}$=-2i$\beta$, $E_{OBC}^{II}$=0 and $|1/\chi(-i\beta)|$ $\to$ 0. Hence, it is localized on the right end of the chain II with real space eigen function that takes the form,
\begin{align} \label{sq15} \tag{SB15}%\tag{S12}
 \begin{cases}
   \psi_{nB}^{II}(-i\beta)= \Bigg[\left( \frac{-t_{L}^{II}}{t_{2}} \right)^n\Bigg]\Phi_{B1}^{II},  \\
    \psi_{nA}^{II}(-i\beta)=  0.
    \end{cases}
\end{align}

According to  (\ref{sq14}) and (\ref{sq15}), as discussed in the main text, the localization of the topological modes with energies +$i\beta$ and -$i\beta$ remain independent of the $\beta$.

\end{widetext}

\end{document}